\documentclass[12pt]{article}

\usepackage{newtxtext,newtxmath}
\usepackage{graphicx}

\usepackage[letterpaper,margin=1in]{geometry}

\renewenvironment{abstract}
	{\quotation}
	{\endquotation}

\date{}

\makeatletter
\renewcommand{\fnum@figure}{\textbf{Figure \thefigure}}
\renewcommand{\fnum@table}{\textbf{Table \thetable}}
\makeatother

\usepackage{scicite}

\usepackage{url}

\newcommand{\ket}[1]{\vert #1 \rangle}

\newcommand{\bra}[1]{\langle #1 \vert}

\newcommand{\braket}[2]{\langle #1 \vert #2 \rangle}

\newcommand{\ketHF}{\vert\text{HF}\rangle}
\newcommand{\braHF}{\langle\text{HF}\vert}
\newcommand{\ketT}[1]{\left|\text{#1}\right\rangle}

\renewcommand{\theta}{\vartheta}

\usepackage{bm}
\usepackage{xcolor}

\def\scititle{
	Explicitly orthogonal coupled cluster theory for ground and excited state intersections among any number of states
}
\title{\bfseries \boldmath \scititle}

\author{
	Leo Stoll$^{1}$,
	Federico Rossi$^{1}$ and
	Henrik Koch$^{1\ast}$ \and
	\small$^{1}$Department of Chemistry, Norwegian University of Science and Technology, Trondheim 7491, Norway.\and
	\small$^\ast$Corresponding author. Email: henrik.koch@ntnu.no
}

\begin{document} 

% Insert the title and author list
\maketitle

% Abstract, in bold
% There are strict length limits, and not all formats have abstracts.
% Consult the journal instructions to authors for details.
% Do not cite any references in the abstract.
\begin{abstract} \bfseries \boldmath

Conical intersections are responsible for the nonadiabatic relaxation of photoexcited molecules back to the ground state. Simulating these radiationless processes requires a high-quality description of the electronic structure, which coupled cluster theory provides. A severe obstacle, however, is the non-Hermitian structure of the coupled cluster Hamiltonian. For same-symmetry states, it yields unphysical complex energies and intersections with the wrong topology. Furthermore, due to the geometric phase effect and bifurcations of the solutions, coupled cluster theory breaks down near ground state intersections. To address this long-standing problem, we derive explicitly orthogonal coupled cluster theory (XO-CC), which restores the correct intersection topology among any number of states using a closed-form correction. In combination with generalized coupled cluster theory, even ground state intersections can be treated. We demonstrate that the framework gives correct two- and three-state intersections in uracil and malonaldehyde with minimal added computational cost compared to the underlying coupled cluster model. In uracil, XO-CC theory describes the relaxation pathway from S$_2$ to S$_0$, which passes through two separate intersections.

\end{abstract}

\newpage
\section*{Introduction}

Internal conversion between electronic states governs a wide range of photophysical and photochemical processes, and its central role in photochemistry has become clear in recent decades \cite{NonadiabaticEventsandConicalIntersections, NonadiabaticityTheImportanceOfConicalIntersections, atchity1991potential, bernardi1990mechanism}. This radiationless relaxation process takes place at conical intersections \cite{UberDasVerhaltenVonEigenwertenBeiAdiabatischenProzessen}, degeneracies of electronic states at which the electronic and nuclear motion are strongly coupled. After excitation, a molecule often passes through a sequence of intersections on its way back to the ground state. Femtosecond pump-probe experiments can nowadays follow these relaxation pathways directly, even in tabletop setups \cite{FemtosecondTimeResolvedPhotoelectronSpectroscopyofPolyatomicMolecules, zewail2000femtochemistry, smith2020femtosecond, worner2025ultrafast}. Their interpretation relies on simulations that combine a quantum treatment of the nuclei \cite{curchod2018ab, MolecularPhotochemistryRecentDevelopmentsinTheory} with an electronic structure method that describes every intersection along the pathway correctly. The latter is not guaranteed, as different electronic structure methods can give qualitatively different intersections \cite{matsika2021electronic}. 

Coupled cluster theory is an attractive electronic-structure method for dynamics simulations, given the accurate treatment of electron correlation it provides. However, conventional coupled cluster theory produces unphysical complex states and energies in regions where same-symmetry states intersect. The intersections themselves are non-conical and of the wrong dimensionality compared to those of Hermitian theories \cite{StructureOptimizationForExcitedStatesWithCorrelatedSecondORderMethodsCC2ADC2, CrossingConditionsInCoupledClusterTheory, NonadiabaticityTheImportanceOfConicalIntersections, Cancoupledclustertheorytreatconicalintersections}. The origin of these artifacts was pointed out by Hättig \cite{StructureOptimizationForExcitedStatesWithCorrelatedSecondORderMethodsCC2ADC2}, who traced them to the anti-Hermitian part of the coupling between the near-degenerate states. Köhn and Tajti \cite{Cancoupledclustertheorytreatconicalintersections} proposed a correction scheme for the energies, which removed their imaginary components. Later, the crossing conditions that describe two-state intersections in coupled cluster theory were analyzed in detail \cite{CrossingConditionsInCoupledClusterTheory, SCC}, and it was shown that a physical description of same-symmetry intersections is recovered if the intersecting states are orthogonal with respect to a positive-definite metric. Based on this realization, similarity constrained coupled cluster (SCC) theory \cite{SCC, SCCSD, SCC2, kjonstad2023communication, doi:10.1021/acs.jctc.4c00276, angelico2026spectroscopic, kjonstad2024photoinduced} was developed, which removes these artifacts through an additional similarity transformation of the coupled cluster Hamiltonian.

Intersections with the ground state pose an additional challenge for single-reference methods specifically. Near these intersections, the ground state acquires a strong multiconfigurational character, and the intermediate normalization common in single-reference methods hinders the correct description of the geometric phase \cite{berry1984quantal,GPECC}. Additionally, near-zero excitation energies lead to bifurcations in the solutions of the CC ground state equations. Other affected methods include time-dependent Hartree-Fock (TDHF) \cite{MEST} and density functional theory (TDDFT) \cite{casida1995time}, algebraic diagrammatic construction (ADC) \cite{PhysRevA.26.2395, https://doi.org/10.1002/wcms.1206}. The consequences include discontinuous potential energy surfaces (PESs), artificial intersections and negative or complex excitation energies \cite{GPECC, tuna2015assessment}. Dynamics simulations with ADC(2) \cite{10.1021/ct4011079} or with similarity constrained coupled cluster singles and doubles (SCCSD) \cite{SCCSD, kjonstad2024photoinduced} therefore break down as soon as the ground state is approached.

Multireference methods, such as complete active space self-consistent field (CASSCF) \cite{ROOS1980157} and multireference configuration interaction (MRCI) \cite{doi:https://doi.org/10.1002/9781119417774.ch9}, do not suffer from these problems, but they require a carefully chosen active space, whose size is limited by the computational cost. Therefore, several remedies have been proposed within single-reference theory \cite{matsika2021electronic}. Spin-flip methods \cite{Krylov2001,Shao2003} obtain the multiconfigurational ground state as an excitation from a high-spin reference. Their spin contamination can be removed in mixed-reference variants \cite{lee2018eliminating}, although other limitations remain \cite{janoš2026limitations}. In coupled cluster theory, specifically, generalized coupled cluster theory (GCC) \cite{GCC} accounts for the geometric phase effect and removes the discontinuities. This framework does not, however, remove the complex energies and intersections of incorrect dimensionality that result from the non-Hermiticity of the coupled cluster Hamiltonian. These remaining artifacts were recently addressed with convex similarity constrained coupled cluster theory (CVX-SCC) \cite{stoll2026cvxscc}. However, the combination of SCC and GCC theory required a careful tailoring of the SCC parameterization to the underlying GCC method. More generally, SCC theory may have multiple solutions, which can lead to discontinuous jumps in energies and properties if not handled carefully, and the optimal form of the SCC wave function parameterization is ambiguous. This implies that the generalization to other levels of coupled cluster approximation is not entirely straightforward \cite{stoll2026cvxscc}.

In this work, we propose explicitly orthogonal coupled cluster theory (XO-CC), a new approach to recover the physically correct description of intersections between any number of chosen electronic states (constrained states). In contrast to SCC theory, the XO-CC formalism leaves the unconstrained states unchanged, always has a unique closed-form solution, and adds only negligible computational cost to the underlying CC method. The XO-CC framework can also be applied, without modifications, to GCC theory, in which case the constrained states may also include the ground state. 

The XO-CC framework can be viewed as a rigorous and general realization of the ideas presented by Köhn and Tajti \cite{Cancoupledclustertheorytreatconicalintersections} and Williams \textit{et al.} \cite{10.1063/5.0151856}, who proposed to modify the anti-Hermitian part of the Hamiltonian to remedy the unphysical description of same-symmetry conical intersections, and of the orthogonality condition of Kjønstad \textit{et al.} \cite{CrossingConditionsInCoupledClusterTheory}. The novel key idea of XO-CC theory is to remove the anti-symmetric part of the Hamiltonian in an $M$-orthonormal basis of the real spectral subspace associated with the intersecting states, and to incorporate this correction into the full-space Hamiltonian. In this way, strictly real states and their properties can be obtained for any number of constrained states. We implement the framework for CC2 \cite{CC2} and CCSD \cite{CCSD}, both with and without GCC, and obtain the XO-CC2, XO-GCC2, XO-CCSD and XO-GCCSD models. We demonstrate that these models produce two-state and multistate intersections with the correct topology and topography. In uracil, XO-GCC with multiple constrained states describes both the S$_1$/S$_2$ and the S$_0$/S$_1$ intersections and the region connecting them. In malonaldehyde, XO-GCC2 correctly describes a three-state intersection including the ground state. Furthermore, vertical excitation energies and oscillator strengths at ground state equilibrium geometries indicate minimal changes to the states of the underlying CC method outside of the intersection regions.

\section*{Results}\label{sec:results}
We first provide a brief presentation of the XO-CC framework and its key properties, and proceed to verify the theory using a modified development branch of the electronic structure program eT \cite{eT}. A detailed derivation of the theory is given in the Materials and methods section.

\subsection*{The XO-CC model}
In equation-of-motion coupled cluster theory, the excited states are described by eigenvectors of the similarity transformed Hamiltonian matrix $\bar{\mathbf{H}}$ in the full manifold of determinants, which is real but non-Hermitian. Near intersections, two eigenstates of the same symmetry can therefore become parallel or form a complex pair. This causes complex energies and wrong intersection dimensionality in standard coupled cluster theory \cite{CrossingConditionsInCoupledClusterTheory, Cancoupledclustertheorytreatconicalintersections, 10.1063/5.0151856}. A physical description is recovered if the intersecting states, the constrained states, are made orthogonal with respect to a positive-definite metric $M$ \cite{CrossingConditionsInCoupledClusterTheory, SCC}. In SCC theory, $M$-orthogonality is achieved indirectly, through an additional similarity transformation of the CC Hamiltonian. On the other hand, XO-CC theory enforces $M$-orthogonality directly, by adding an explicit closed-form correction to the CC Hamiltonian matrix.

Within the real spectral subspace $V$ (of the full manifold of determinants) associated with the $m$ constrained states, $\bar{H}$ is represented by an $m\times m$ matrix $\bar{\mathbf{H}}^V$. In an $M$-orthonormal basis of this subspace, the eigenvectors of $\bar{\mathbf{H}}^V$ are real and $M$-orthogonal exactly when this matrix is symmetric. Its anti-symmetric part is thus what prevents $M$-orthogonality. The XO-CC correction, written as $\zeta \mathbf{C}$, is constructed such that it removes this anti-symmetric component, that is
\begin{equation}
    \tilde{\mathbf{H}} = \bar{\mathbf{H}} - \zeta\mathbf{C}, \qquad
    \tilde{\mathbf{H}}^V = \mathbf{S},
\end{equation}
where $\tilde{\mathbf{H}}$ ($\tilde{\mathbf{H}}^V$) is the XO-CC Hamiltonian matrix in the full manifold (in the $M$-orthonormal basis of the subspace), $\mathbf{S}$ is the symmetric part of $\bar{\mathbf{H}}^V$, $\zeta$ is the total anti-symmetric coupling and $\mathbf{C}$ is the normalized correction matrix. The eigenvectors of $\tilde{\mathbf{H}}^V$ are thus $M$-orthogonal. The real spectral subspace $V$ is invariant under the XO-CC Hamiltonian, and, consequently, its eigenvectors in this subspace are exact eigenvectors also in the full manifold. Thus, the full space XO-CC Hamiltonian produces $M$-orthogonal constrained states. These states are then guaranteed to have real energies, and their intersections are conical with correct dimensionality \cite{CrossingConditionsInCoupledClusterTheory, SCC}. The restoration of the correct crossing conditions can also be understood as the direct consequence of the constrained states being the eigenvectors of a real symmetric subspace Hamiltonian matrix. Even for parallel states and complex pairs, the real spectral subspace $V$ remains well defined and smooth \cite{LINALG, KATO}, allowing a continuous definition of XO-CC theory. The only requirements are that the set of constrained states must be energetically separated from the other eigenstates of the CC Hamiltonian, and that both parts of a complex pair must be constrained together. These two requirements determine how the constrained states must be selected near intersections.

The correction $\zeta \mathbf{C}$ is given in closed form by the left and right eigenvectors and the associated energies of the underlying method. In principle, no additional equations have to be solved iteratively. The correction is the unique smallest change that symmetrizes $\bar{\mathbf{H}}^V$, and it leaves all other states unchanged. The average energy of the constrained states is preserved, and only minimal computational cost is added to the underlying method. The total anti-symmetric coupling $\zeta$ vanishes in the FCI limit when we use the natural metric \cite{SCCSD} as $M$. This metric is used throughout this work.

To see how the XO-CC correction acts on the individual states, it is convenient to express the CC Hamiltonian in the basis of the XO-CC constrained states themselves, which form a specific $M$-orthonormal basis of $V$. In this basis, the symmetric part of the CC Hamiltonian is diagonal and contains the XO-CC energies, and the states are coupled only by pairwise anti-symmetric couplings $\zeta_{ij}$. These couplings have a direct interpretation: they indicate how much each pair contributes to the XO-CC energy correction (see Materials and methods). The total anti-symmetric coupling $\zeta$ is related to the pairwise couplings as $\zeta = (\sum_{i<j}\zeta_{ij}^2)^{1/2}$. For two constrained states the XO-CC energies $\tilde{E}_k$ are related to the CC energies $E_k$ by
\begin{equation}
    (\tilde{E}_l - \tilde{E}_k)^2 = (E_l - E_k)^2 + 4\zeta^2.
\end{equation}
That is, the XO-CC constrained states are pushed apart symmetrically about their mean, and complex CC pairs, for which $(E_l - E_k)^2 < 0$, become real. An XO-CC intersection therefore occurs on or within the intersection hypertube \cite{CrossingConditionsInCoupledClusterTheory, 10.1063/5.0151856} of the underlying method. When more than two states are constrained, the pairwise anti-symmetric couplings $\zeta_{ij}$ connect every pair of them. For energetically well-separated states, however, a pairwise coupling has little effect, as its contribution to the energy correction is inversely proportional to the energy gap between the two states (see Eq.~\eqref{eq:xo_second_order}). Close to a two-state intersection, the other constrained states therefore barely alter the correction to the intersecting pair, which remains nearly the same as when only these two states are constrained. The XO-CC correction is therefore local in energy: even with multiple constrained states it acts strongly only on near-degenerate states.

With the XO correction applied to GCC theory \cite{GCC}, which removes bifurcations near ground state intersections and correctly describes the geometric phase of the wave function, the ground state can also be constrained. We refer to the resulting methods as XO-GCC methods. Finally, if HF shows discontinuities or nonanalytical points due to bifurcations in the HF solutions, a CVX-HF \cite{CVXHF} reference may be used instead. 

\subsection*{Two-state intersections in uracil}

\begin{figure}[pbt!]
    \centering
    \includegraphics[width=\linewidth]{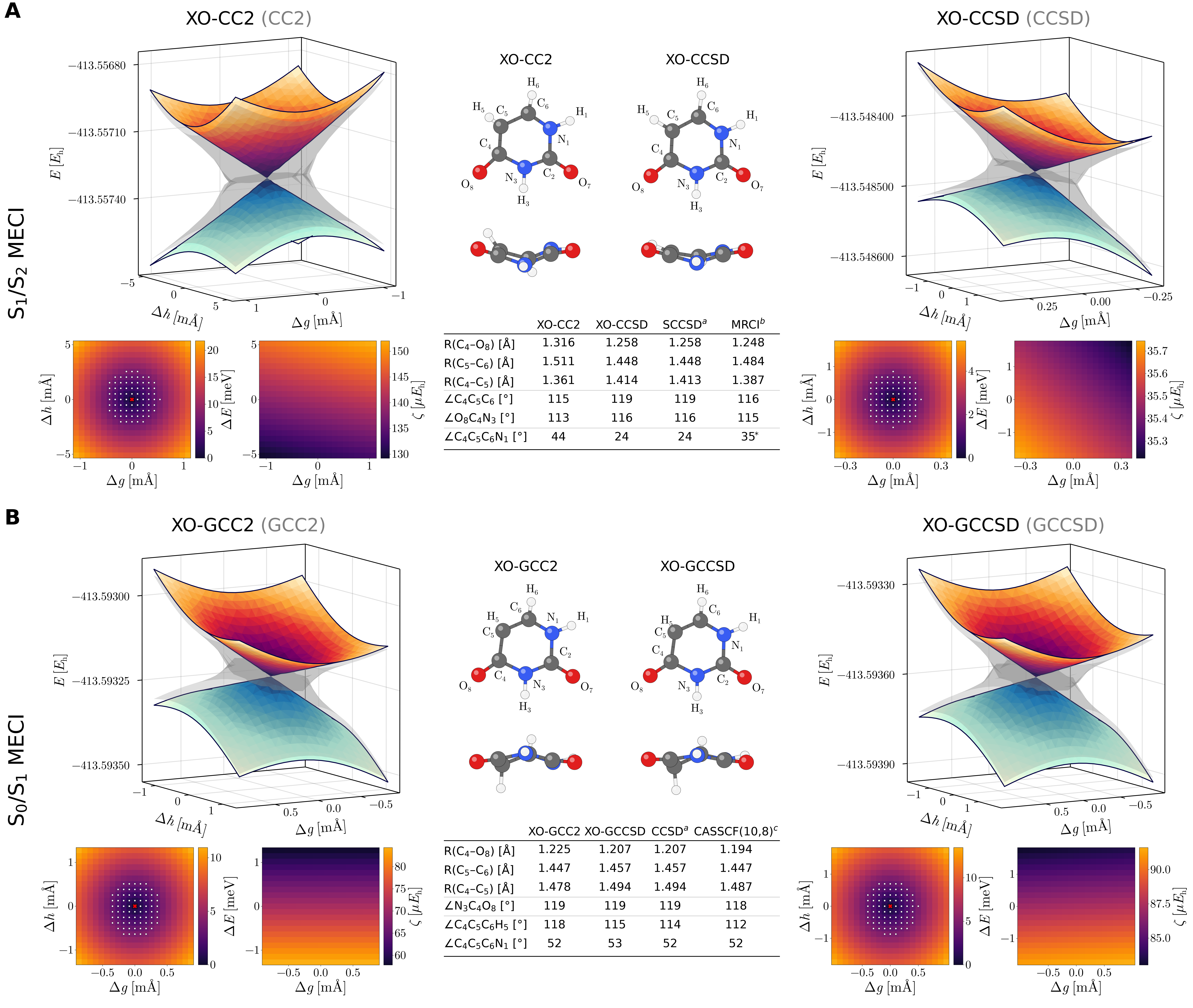}
    
    \footnotesize
    $^\text{a}$ $\varepsilon$-MECI from Angelico \textit{et al.} \cite{angelico2025determining} \,
    $^\text{b}$ Values from Matsika \cite{10.1021/jp048284n} \,
    $^\text{c}$ Values from Nachtigallov\'a \textit{et al.} \cite{10.1021/jp201327w} \\
    $^{*}$ Extrapolated from geometry data.
    
    \caption{\footnotesize \textbf{XO-CC minimum energy conical intersections (MECIs) in uracil, obtained with the cc-pVDZ basis.} In the 2D scans, the equations are converged to a residual threshold of $10^{-8}$ a.u.. The MECIs were determined using the tube algorithm \cite{angelico2025determining} with a finite difference approximation for the required molecular gradients. The branching planes at the MECI geometries were similarly obtained numerically by a finite difference approach and a singular value decomposition of the lifting of the degeneracy, see Supplementary Materials for details. XO-CC/XO-GCC PESs (colored) are overlaid with those of CC/GCC (gray) in the same plane. The GCC models were run with one projected state. Heatmaps of the energy differences between the intersecting states have complex CC/GCC points indicated with white dots, and $\zeta$ is the anti-symmetric coupling. A: The molecular geometries of the S$_1$/S$_2$ MECIs of XO-CC2 and XO-CCSD, and results from a 2D scan in their respective branching planes. B: The molecular geometries of the S$_0$/S$_1$ MECIs ($^6$S$_5$) of XO-GCC2 and XO-GCCSD, and results from a 2D scan in their respective branching planes.} \label{fig:Uracil}
\end{figure}

The photostability of nucleobases is crucial to the resistance of DNA to UV-induced damage. This stability is related to a high density of conical intersections which efficiently take the excited molecule back to the ground state. Here, we will consider conical intersections in the nucleobase uracil, which have been studied extensively in the literature \cite{10.1021/jp048284n, NonadiabaticEventsandConicalIntersections, 10.1021/jp201327w, 10.1021/jp0723665, angelico2025determining}. In Figure \ref{fig:Uracil} A, we show 2D scans in the branching planes of XO-CC S$_1$/S$_2$ minimum energy conical intersections (MECIs). The CC2 and CCSD results display the expected complex excitation energies, intersection ovals of wrong topology, and a nonlinear lifting of the degeneracy. On the other hand, by enforcing the same crossing conditions as in Hermitian theories, XO-CC theory recovers conical intersections with the correct characteristics. Additionally, we observe that the corrected intersections lie within the intersection oval of the underlying CC method, as expected. In Figure \ref{fig:Uracil} B, we show the corresponding 2D scans for XO-GCC S$_0$/S$_1$ ground state MECIs. As previously reported, GCC2 and GCCSD produce continuous PESs, and recover the correct ordering of the electronic states near the ground state intersection \cite{GCC, stoll2026cvxscc}. However, GCC theory does not correct the numerical artifacts caused by the anti-symmetric component of the coupled cluster Hamiltonian, which causes the same numerical artifacts as for excited state intersections in standard coupled cluster theory. This limitation of GCC theory has previously been rectified with CVX-SCC theory \cite{stoll2026cvxscc}. Our results demonstrate that XO-GCC theory is able to achieve similar results with a closed-form correction, avoiding the numerical solution of additional nonlinear equations.

Key bond lengths and angles of the MECI structures are provided in Figure \ref{fig:Uracil}. While the pathways for internal conversion in uracil deserve a more detailed study, here, we will only give a brief account of the MECIs. The XO-CCSD and XO-GCCSD MECIs lie very close to those of SCCSD and CCSD ($\varepsilon$-MECI) \cite{angelico2025determining}, respectively. On the other hand, the XO-CC2 and XO-GCC2 MECIs show larger deviations. This is expected as the underlying CC approximation is different. Still, all XO-CC MECIs show the same distinctive features as the multireference MECIs. The S$_1$/S$_2$ MECIs display the characteristic boat-shaped ring \cite{10.1021/jp0723665, angelico2025determining, 10.1021/jp048284n}, with the strength of the ring bend being largest in XO-CC2 ($\angle$C$_4$C$_5$C$_6$N$_1$ = $44^\circ$) and smallest in XO-CCSD ($24^\circ$), and MRCI \cite{10.1021/jp048284n} lying in between ($35^\circ$). Similarly, the S$_0$/S$_1$ MECIs (labeled $^6$S$_5$ \cite{angelico2025determining}) display ring puckering and an out-of-plane hydrogen (H$_5$) which is also observed in CASSCF(10,8) \cite{10.1021/jp201327w}. In fact, all reported methods agree on the ground state MECI, with bond lengths differing up to only 0.03~\AA\ and bond angles up to $1^\circ$ (dihedral angles $6^\circ$). The MECIs were obtained with the tube algorithm developed by Angelico \textit{et al.} \cite{angelico2025determining}, which can be run with a finite difference scheme for the required molecular gradients. We refer to the Supplementary Materials for details.

\subsection*{Relaxation pathway in uracil}

\begin{figure} [p!]
    \centering
    \includegraphics[width=\linewidth]{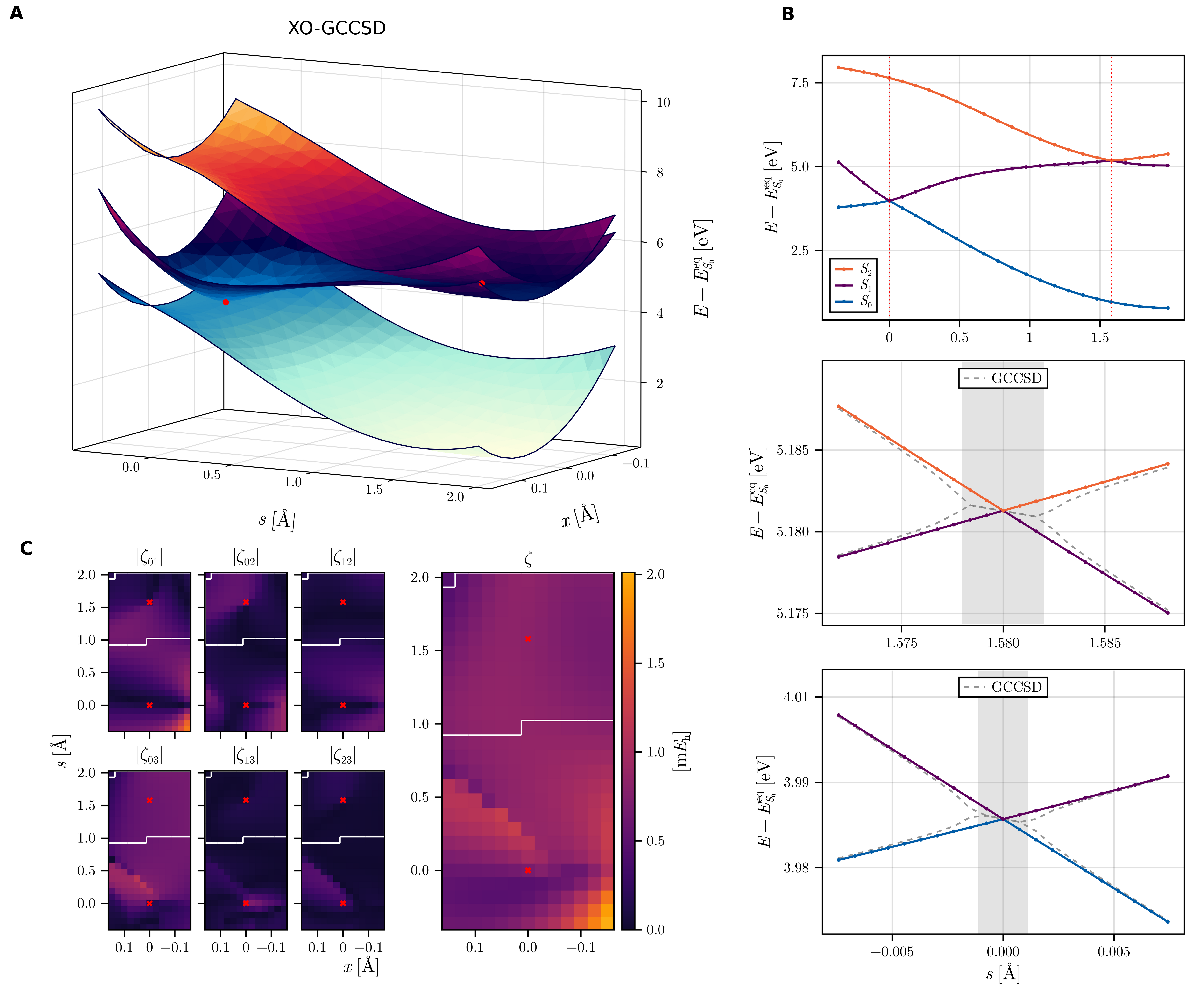}
    \caption{\footnotesize \textbf{Uracil XO-GCCSD 2D scan through internal coordinate space, passing through S$_0$/S$_1$ and S$_1$/S$_2$ same-symmetry conical intersections.} The GCCSD calculations were run with an adaptive number of projected states: as a baseline one state was projected, but close to the S$_1$/S$_2$ intersection both states were projected instead (condition $|E_2 - E_1| < 1$ eV). The ground state and three lowest excited states were constrained in the XO correction. The S$_3$ state was included among the constrained states as it crosses S$_2$ within the 2D-plane. Equations were converged to $10^{-8}$ a.u. and a cc-pVDZ basis was used in all calculations. The intersection points are marked in red. Energies are relative to the XO-GCCSD (one projected state and four constrained states) energy at the CCSD/cc-pVDZ ground state equilibrium geometry. A: PES of the lowest three electronic states. B: Cross sections at $x = 0$ of the PESs of the 2D scan (real components for GCCSD). Shading indicates regions where GCCSD has complex energies. The plots show GCCSD with one projected state at the S$_0$/S$_1$ intersection, and two projected states at the S$_1$/S$_2$ intersection. C: Pairwise anti-symmetric couplings $|\zeta_{ij}|$ of the constrained states, and the total anti-symmetric coupling $\zeta$. The region where two states were projected in GCCSD is indicated by a white outline. The discontinuities in the couplings involving S$_3$ stem from an additional crossing of S$_3$ and the unconstrained S$_4$, where S$_3$ changes character.}
    \label{fig:Uracil_threestate}
\end{figure}

With SCC theory \cite{SCC, SCCSD} and CVX-SCC theory \cite{stoll2026cvxscc}, excited state intersections and ground state intersections can already be described, but only separately, each with their own SCC wave function parameterization. In contrast, the XO-CC correction with four constrained states (S$_0$-S$_3$) can be applied to an underlying GCCSD calculation in a region involving conical intersections between several different states. In Figure \ref{fig:Uracil_threestate}, we explore a plane that cuts through S$_0$/S$_1$ and S$_1$/S$_2$ intersections. Here, the XO-CC correction with four constrained states applied to GCCSD describes both intersections and the region connecting them. This indicates that a single form of the XO-CC correction can cover all parts of the potential energy landscape required to model internal conversion from S$_2$ to the ground state. However, an adaptive scheme was used for the underlying GCCSD model. As the S$_1$/S$_2$ intersection is approached, the projection of S$_1$ alone becomes unstable, and both states are projected instead. We stress that the necessity of this adaptive scheme is due to the underlying GCCSD model, not due to the XO-CC correction. The discontinuities due to the transition between one and two projected states in the underlying GCCSD calculation are small ($<1.5$ meV, see the Supplementary Materials).

In the 2D scan, the total anti-symmetric coupling (see Figure~\ref{fig:Uracil_threestate}~C) is discontinuous along a line passing between the S$_0$/S$_1$ and S$_1$/S$_2$ intersections. This discontinuity is caused by an intersection of the constrained S$_3$ state with the unconstrained S$_4$ state, where S$_3$ changes character. As a consequence, the anti-symmetric couplings involving S$_3$ are discontinuous in this region. Since the effect of a pairwise coupling on the XO-CC energies decreases with the energy difference of the two states, the discontinuities shift the energies of the lower-lying states S$_0$-S$_2$ only marginally ($<0.5$ meV, see Supplementary Materials). The scan thus demonstrates that the locality of the XO-CC correction alleviates the main limitation of the framework, the requirement that the constrained states must be energetically separated from the unconstrained ones for a smooth XO correction to be obtained. In practice, we observe that a crossing involving the highest constrained state leaves the lower constrained states essentially unaffected.

In Figure \ref{fig:Uracil_threestate} B, we show 1D cuts through the two-state intersections. In these cuts, the locality of the XO-CC correction can be observed directly. The changes to the GCC energies are largest at the intersections, and become smaller moving away from them. In the 2D scan of Figure \ref{fig:Uracil_threestate} A, the bare GCCSD results are not shown for this reason: at the displayed scale, they are too close to the XO-GCCSD results for any difference to be observed.

\subsection*{Vertical excitation energies and oscillator strengths}

\begin{table}[p!]
\centering
\footnotesize
\begin{tabular}{lrrrrrr}
\hline
Method & $E_0$ [$E_\mathrm{h}$] & \multicolumn{5}{c}{$\omega_n$ [eV] / $f_n$} \\ \cline{3-7}
 & & $n=1$ & $n=2$ & $n=3$ & $n=4$ & $n=5$ \\ \hline
\multicolumn{7}{l}{\textit{HOF}, $2\,^1A''$/$3\,^1A''$ (k=3, l=4)} \\

CC2$^{(3,5)}$ & $-$175.191996 & 5.235146 & 7.183674
& $\mathbf{7.391754}$ & 9.356358 & $\mathbf{9.491537}$ \\
 & & 0.0009 & 0.0064 & $\mathbf{0.0111}$ & 0.0108 & $\mathbf{0.0003}$ \\[2pt]
XO-CC2$^{(3,5)}$ & $-$175.191996 & 5.235146 & 7.183674
& $\mathbf{7.391747}$ & 9.356358 & $\mathbf{9.491544}$ \\
 & & 0.0009 & 0.0064 & $\mathbf{0.0111}$ & 0.0108 & $\mathbf{0.0003}$ \\[2pt]
CCSD & $-$175.194048 & 5.197586 & 7.138723
& $\mathbf{7.871465}$ & $\mathbf{9.188986}$ & 10.067266 \\
 & & 0.0006 & 0.0050 & $\mathbf{0.0125}$ & $\mathbf{0.0003}$ & 0.0136 \\[2pt]
XO-CCSD & $-$175.194048 & 5.197586 & 7.138723
& $\mathbf{7.871460}$ & $\mathbf{9.188991}$ & 10.067266 \\
 & & 0.0006 & 0.0050 & $\mathbf{0.0125}$ & $\mathbf{0.0003}$ & 0.0136 \\[2pt]

\multicolumn{7}{l}{\textit{PSB3}, $1\,^1A'$/$2\,^1A'$ (k=0, l=1)} \\

GCC2 & $\mathbf{-249.090121}$
& $\mathbf{4.239809}$ & 6.530744 & 6.543306 & 6.942071 & 7.183791 \\
 & & $\mathbf{0.7394}$ & 0.1858 & 0.0004 & 0.2537 & 0.0015 \\[2pt]

XO-GCC2 & $\mathbf{-249.090124}$
& $\mathbf{4.239977}$ & 6.530828 & 6.543390 & 6.942155 & 7.183875 \\
 & & $\mathbf{0.7400}$ & 0.1855 & 0.0004 & 0.2536 & 0.0015 \\[2pt]

GCCSD & $\mathbf{-249.132296}$
& $\mathbf{4.389195}$ & 6.619434 & 6.625467 & 7.042978 & 7.296238 \\
 & & $\mathbf{0.7047}$ & 0.1461 & 0.0005 & 0.1949 & 0.0013 \\[2pt]

XO-GCCSD & $\mathbf{-249.132301}$
& $\mathbf{4.389495}$ & 6.619584 & 6.625617 & 7.043128 & 7.296388 \\
 & & $\mathbf{0.7040}$ & 0.1463 & 0.0005 & 0.1949 & 0.0013 \\[2pt]
\hline
\end{tabular}

\caption{\footnotesize \textbf{Excitation energies and oscillator strengths of the five lowest singlet excited states at the ground state equilibrium geometries of HOF and PSB3.} The geometries were optimized with CCSD/cc-pVTZ using eT \cite{eT}. All energies are computed with the aug-cc-pVDZ basis and with a convergence threshold of $10^{-8}$ $E_h$, and $E_0$ is the ground state energy. For each method, the first line gives the excitation energies $\omega_n$ and the second line the EOM-CC oscillator strengths $f_n$ for the transitions from the ground state to state $n$. The states $n$ are numbered in order of energy for each method. The constrained pair $k/l$ is given for each molecule, the labels $k$ and $l$ follow the CCSD ordering. The superscript $(3,5)$ indicates that the HOF states $k$ and $l$ correspond to CC2 roots 3 and 5. Values for the constrained states are bold. For PSB3, since the ground state is among the constrained states, all excitation energies are shifted and all oscillator strengths may change in the XO step. The GCC calculations use one projected state.} \label{tab:docc_vertical_short}
\end{table}

Away from intersections, the XO-CC correction should leave the description of the underlying coupled cluster model essentially unchanged. This is expected from the theory: the XO-CC correction predominantly affects constrained states that are close in energy, whereas well separated states change only weakly. To verify this, Table \ref{tab:docc_vertical_short} compares ground state energies, vertical excitation energies and oscillator strengths of the XO-CC models and their underlying models at the ground state equilibrium geometries of HOF and the protonated Schiff base model system (PSB3). In each molecule, a pair of same-symmetry states is constrained: the $2\,^1A''$ and $3\,^1A''$ excited states in HOF, and the ground state and the bright $2\,^1A'$ state in PSB3, which requires XO-GCC models.

In HOF, the ground state and all unconstrained excited states are identical to those of CC2 and CCSD to all reported digits, and so are all oscillator strengths, including those of the constrained states. Only the energies of the two constrained states change, and they are shifted symmetrically about their mean, by $\pm 7.0~\mu$eV for XO-CC2 and $\pm 4.8~\mu$eV for XO-CCSD.

In PSB3, the ground state is itself a constrained state. The XO-GCC2 and XO-GCCSD models lower the ground state energy by 3.1 and 5.5~$\mu E_\mathrm{h}$, respectively, and raise the energy of the $2\,^1A'$ state by the same amount. Because excitation energies are measured relative to the ground state, the excitation energies of the unconstrained states also increase. The increase is uniform, by 0.08 and 0.15~meV for the two methods, respectively. This uniform increase equals the lowering of the ground state (within precision of the table), that is, the absolute energies of the unconstrained states are unchanged. The oscillator strengths change for all transitions, as the transitions now originate from the modified ground state. However, the changes are below 0.2\,\% for all bright transitions (oscillator strengths $f_n > 0.1$).

In summary, these results demonstrate that the XO-CC correction is negligible away from intersections between the constrained states, both for excited state pairs and for pairs involving the ground state. The XO-CC energy shifts are significantly below the typical error ranges for excitation energies of the CC2 and CCSD methods, which are in the range of 0.1-0.3 eV \cite{loos2020mountaineering}.

\subsection*{Malonaldehyde three-state intersection}

\begin{figure} [p!]
    \centering
    \includegraphics[width=\linewidth]{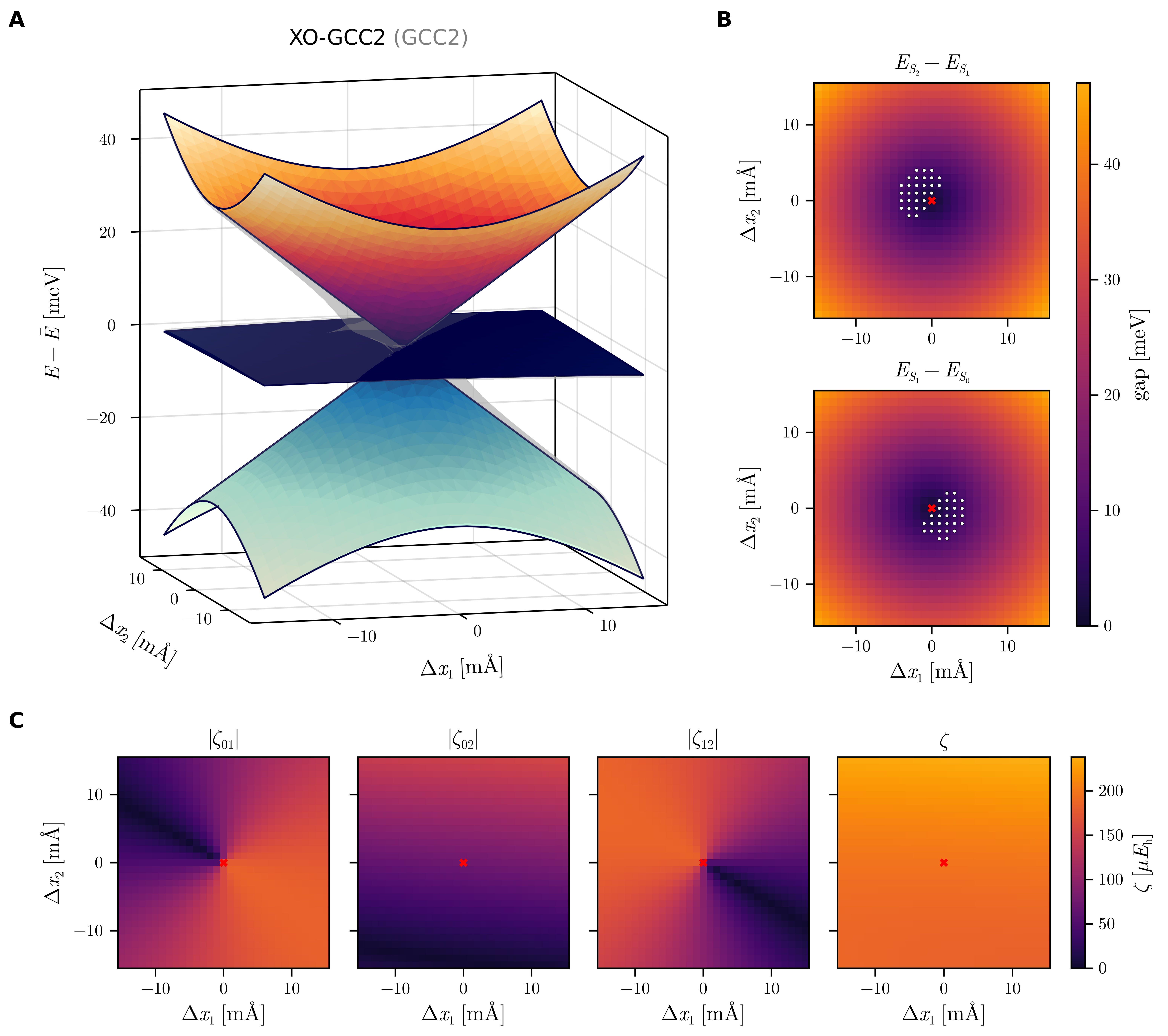}
    \caption{\footnotesize \textbf{XO-GCC2 (colored) and GCC2 (gray) 2D scan of a malonaldehyde S$_0$/S$_1$/S$_2$ three-state same-symmetry ground state intersection using a 6-31G* basis.} The GCC calculations used two projected states, and all three states were constrained in the XO step. Equations were converged to a threshold of $10^{-7}$ a.u.. The intersection is located starting from the CASSCF three-state intersection of Coe and Martínez \cite{coe2005competitive}, in the space spanned by its g- and h-vectors. Details on the intersection and the 2D-plane of the scan are provided in the Supplementary Materials. A: The PESs of the three intersecting states. B: Pairwise energy differences between the states. The intersection point is marked by a red cross. The white dots indicate complex energy pairs for the GCC states. C: Pairwise anti-symmetric couplings $|\zeta_{ij}|$ between the constrained states, and total anti-symmetric coupling $\zeta$, in the plane.}
    \label{fig:malonaldehyde}
\end{figure}

The 2D scan through two conical intersections in uracil (see Figure \ref{fig:Uracil_threestate}) confirms that XO-CC theory with several constrained states can describe pairwise intersections within a set of more than two constrained states. However, XO-CC theory is also able to describe regions where more than two of the constrained states intersect at once. In Hermitian theories, a same-symmetry intersection of three states requires five conditions to be fulfilled: two energy differences and three couplings must vanish \cite{coe2005competitive}. These intersections therefore form seams of dimension $N-5$, where $N$ is the number of internal nuclear coordinates. Since the XO-CC Hamiltonian is symmetric in the $M$-orthonormal basis of the subspace associated with the constrained states, and the constrained states are the eigenvectors of this symmetric Hamiltonian, the same five conditions apply to XO-CC three-state intersections. In Figure \ref{fig:malonaldehyde} we show a 2D scan through an XO-GCC2 three-state intersection in malonaldehyde. In parts of the intersection region, two of the GCC2 states form a complex pair, either S$_0$/S$_1$ or S$_1$/S$_2$, while the third stays real. On the other hand, the XO-GCC2 PESs are real on the entire scan, smooth across the boundaries of the GCC2 complex regions, and preserve the energy average of the three GCC2 states. All three pairwise energy differences vanish at the intersection point and increase linearly away from it. The total anti-symmetric coupling $\zeta$ is smooth on the entire scan.

\section*{Discussion}

Explicitly orthogonal coupled cluster theory produces two-state and multistate intersections with the same properties as those of Hermitian theories. It achieves this by adding a minimal correction to the coupled cluster Hamiltonian matrix, which modifies a specified set of constrained states, and leaves other states and their properties unchanged. Away from intersections, the XO-CC correction is much smaller than the error ranges of the underlying methods. 

The XO-CC correction is explicitly defined, unique, and is determined with closed-form expressions. With specific choices of metric, either the FCI limit or correct size-scaling properties can be enforced. While SCC theory is already able to treat two-state excited state intersections within a coupled cluster framework, the best choice of wave function parameterization (choice of additional similarity transformations) to employ is not entirely clear, and the framework does not enforce unique corrections to the underlying CC method. Further, SCC theory has to be specifically tailored to any underlying CC method it aims to correct. For example, significant changes had to be made to the parameterization to allow the description of ground state intersections with CVX-SCC theory. In contrast, the XO-CC correction can be directly applied to any CC method, including GCC methods, without modifications of the parameterization. The resulting XO-GCC models allow a computationally less demanding description of ground state intersections compared to CVX-SCC theory, and remove the possibility of multiple solutions in the $M$-orthogonalization procedure.

The XO-CC construction makes the extension to an arbitrary number of constrained states trivial. With more than two constrained states, the framework can describe regions connecting a series of conical intersections among any of the constrained states. This makes XO-CC theory particularly suitable for dynamics simulations of internal conversion through a series of conical intersections, and all the way to the ground state. Moreover, XO-CC models with several constrained states can also describe multistate intersections between more than two constrained states in a single point. 

The main limitation of the XO-CC formulation is the requirement that the constrained states must be energetically separated from other states in order to obtain a smooth XO-CC Hamiltonian. As higher-lying electronic states are closer in energy, and cross more frequently, this condition is expected to be broken in practical applications. However, as the pairwise effect of the XO-CC correction decreases with energetic separation of the states, higher-lying state crossings do not significantly affect the lower constrained states. This is what we observe in our calculations on the relaxation pathway in uracil. The locality of the correction also implies that a potential adaptive scheme, where different states are constrained in different regions of internal coordinate space, should not suffer from significant discontinuities. As the XO-CC correction is computationally cheap, another approach to XO-CC dynamics is to simply constrain all relevant states throughout the simulation. The next steps are to derive and implement analytical gradients and nonadiabatic coupling vectors for GCC theory and the XO correction step, and to apply the XO-CC framework in nonadiabatic dynamics simulations. This work is in progress, and we will report on the results in future work.

\section*{Materials and methods}

\subsection*{Coupled cluster theory}
The coupled cluster ground state is obtained by applying the exponential of the cluster operator to a reference determinant, in this work the closed-shell Hartree-Fock (HF) state $\ketT{HF}$,
\begin{equation}
    \ketT{CC} = \exp(T) \ketHF.
\end{equation}
The cluster operator is a sum of operators of increasing excitation rank,
\begin{equation}
    T = \sum_{i=1}^{n_\text{cc}} T_i,
\end{equation}
\begin{equation}
    T_i = \sum_{\mu_i} t_{\mu_i} \tau_{\mu_i},
\end{equation}
where $\tau_{\mu_i}$ creates the $i$-fold excited determinant $\ket{\mu_i} \equiv \tau_{\mu_i} \ketHF$ and $t_{\mu_i}$ is the associated amplitude. The CC truncation is defined by the highest rank $n_\text{cc}$ retained in $T$. If no truncation is made, the resulting full coupled cluster (FCC) wave function is identical to the full configuration interaction (FCI) wave function, and they have the same energy. In practice, as the computational scaling increases rapidly with excitation rank, strongly truncated coupled cluster models are applied. For instance, the cost of the CCSD model ($n_\text{cc} = 2$) \cite{CCSD} scales as $N^6$ with the number of molecular orbitals $N$. The CC2 model \cite{CC2} approximates the doubles equations perturbatively, and achieves a lower scaling of $N^5$.

The amplitudes are obtained by projecting the Schr\"odinger equation for the similarity transformed Hamiltonian $\bar{H} \equiv \exp(-T)H\exp(T)$, where $H$ is the Born-Oppenheimer electronic Hamiltonian \cite{MEST}, onto the projection manifold $\{ \ket{\mu_i} | 0 < i \leq n_\text{cc} \}$,
\begin{equation}
    \Omega_{\mu_i} \equiv\bra{\mu_i} \bar{H} \ketHF = 0,
\end{equation}
and the ground state energy from
\begin{equation}\label{eq:CC_energy}
    E_\text{CC} = \braHF \bar{H} \ketHF.
\end{equation}
In the following, $\ket{\mu}$ denotes a general determinant in the projection manifold, and $\ket{\mu_0}$ refers to the reference state. We will work with spin-adapted singlet models, where we restrict the excitation operators $\tau_i$ to singlet excitations. Correspondingly, we will let $\{\bra{\mu}\}$ denote the biorthonormal bra basis ($\braket{\mu}{\nu} = \delta_{\mu\nu}$) rather than the adjoint one \cite{MEST}.

Excited states are described with equation-of-motion coupled cluster (EOM-CC) theory \cite{EOM-CC}, where the right and left states are expanded in the full manifold $\{ \ket{\mu_i} | 0 \leq i \leq n_\text{cc} \}$,
\begin{equation}
\begin{aligned}
    \ket{R^k} &= \sum_\nu \ket{\nu} r_{\nu}^k \\
    \bra{L^k} &= \sum_\mu \bra{\mu} l_{\mu}^k.
\end{aligned}
\end{equation}
Their amplitudes are the right and left eigenvectors of the matrix representation of $\bar{H}$,
\begin{equation}
    \mathbf{\bar{H}} = 
    \begin{pmatrix} 
        E_\text{CC} & \bm{\eta}^T \\
        \bm{0} & \mathbf{A} + E_\text{CC} \bm{I}
    \end{pmatrix},
\end{equation}
with the CC Jacobian \cite{MEST}
\begin{equation}
    A_{\mu\nu} = \bra{\mu} [\bar{H}, \tau_\nu] \ketHF
\end{equation}
and
\begin{equation}
    \eta_\nu = \braHF [\bar{H}, \tau_\nu] \ketHF.
\end{equation}
As $\mathbf{\bar{H}}$ is block triangular, the excitation energies $\omega_k \equiv E_k - E_0$ and the excited-determinant components of the left and right states are determined from the nonsymmetric Jacobian matrix,
\begin{equation}
\begin{aligned}
    \mathbf{A} \bm{r}^k &= \omega_k \bm{r}^k \\
    \mathbf{A}^T \bm{l}^k&= \omega_k \bm{l}^k.
\end{aligned}
\end{equation}
The reference components are
\begin{equation}
\begin{aligned}
    r^k_0 &= \bm{\eta}^T\mathbf{A}^{-1}\bm{r}^k= -\bar{\bm{t}}^T\bm{r}^k\\
    l^k_0 &= 0,
\end{aligned}
\end{equation}
where $\bar{\bm{t}}$ is implicitly defined. The left and right states are neither orthonormal nor adjoint, and are biorthonormalized explicitly ($\braket{L^k}{R^l} = \delta_{kl}$).

\subsection*{Generalized coupled cluster theory}
Close to a ground state conical intersection, the intermediate normalization of the CC wave function is incompatible with the geometric phase \cite{longuet1958studies,berry1984quantal,GPEdiab}. As a result, the CC amplitudes diverge when traveling around a ground state intersection. Additionally, the CC Jacobian may acquire small or negative eigenvalues at the ground state intersection region, which lead to a nonconvex optimization problem for the ground state amplitudes, and multiple solutions appear \cite{GPECC}. Generalized coupled cluster (GCC) theory \cite{GCC} removes these artifacts. In GCC theory, the biorthonormal projectors $P_i = \ket{R^i}\bra{L^i}$ are used to remove the components of a selection of projected states from the projection manifold and the cluster amplitudes. In the remaining manifold, the Jacobian has only positive eigenvalues (positive real parts), such that the amplitude equations define a convex problem \cite{GCC}. To recover the contributions of the removed components, the energies and states are obtained in the full manifold $\{ \ketHF, \ket{\mu} \}$, where the GCC Hamiltonian matrix reads
\begin{equation}
    \mathbf{\bar{H}} = 
    \begin{pmatrix} 
        E_\text{CC}  & \bm{\eta}^T \\
        \bm{\Omega}^{P} & \mathbf{A} + E_\text{CC} \bm{I} + \mathbf{W}
    \end{pmatrix},
\end{equation}
with the leftover terms of the projected components
\begin{equation}
    \Omega_{\mu}^{P} \equiv \sum_{i=1}^n r^i_\mu \bra{L^i}\bar{H}\ketHF, \qquad
    W_{\mu \nu} \equiv \bra{\mu}\tau_\nu \bar{H}\ketHF - E_\text{CC} \, \delta_{\mu \nu} \,.
\end{equation}
The eigenvalue problem
\begin{equation}
    \mathbf{\bar{H}} \bm{x}^k = E_k \bm{x}^k, \hspace{15 pt} k = 0,1,\dots\,
\end{equation}
gives both the GCC ground state and the excited states. In this work we apply the GCCSD model \cite{GCC} and the GCC2 model \cite{stoll2026cvxscc}, which have computational cost scaling as CCSD and CC2, but with larger prefactors.

\subsection*{Explicitly orthogonal coupled cluster theory}
Because the coupled cluster Hamiltonian $\bar{H}$ is not Hermitian, same-symmetry states can turn into complex pairs close to their intersections. The intersections themselves are governed by the CC crossing conditions \cite{CrossingConditionsInCoupledClusterTheory} instead of those of Hermitian theories \cite{TheCrossingOfPotentialSurfaces, UberDasVerhaltenVonEigenwertenBeiAdiabatischenProzessen}, and have the wrong topology and topography.

The correct crossing conditions are recovered if the intersecting states $\ket{R^k}$ and $\ket{R^l}$ are required to be orthogonal with respect to a positive-definite metric $M$ \cite{CrossingConditionsInCoupledClusterTheory},
\begin{equation}
    O(k,l)\equiv\bra{R^k}M\ket{R^l} = 0 \hspace{15 pt} \text{for} \hspace{15 pt} k\neq l,
\end{equation}
referred to as the orthogonality condition or $M$-orthogonality. $M$-orthogonality implies that the states are real, have real eigenvalues, that conical intersections between them are defined by the correct number of crossing conditions and that degeneracies are lifted linearly in their branching plane \cite{CrossingConditionsInCoupledClusterTheory, SCC}. In SCC theory, $M$-orthogonality is enforced by introducing additional flexibility in the wave function with a second similarity transformation of the effective CC Hamiltonian, and optimizing the SCC wave function parameters with the constraint given by the orthogonality condition \cite{SCC, SCCSD, SCC2}. This leads to an \textit{ab initio} theory, which produces conical intersections qualitatively similar to those of Hermitian theories. However, the exact choice of the additional similarity transformation is ambiguous, and the solution of the coupled nonlinear SCC set of equations adds additional computational cost. Here, we propose a more direct way of enforcing $M$-orthogonality.

In XO-CC theory, the orthogonality condition is enforced by introducing the minimal correction to the CC effective Hamiltonian matrix that will cause a selection of constrained states to become $M$-orthogonal. Initially, we will assume two constrained states. The extension to more states follows below. We consider two linearly independent and real right eigenvectors $\mathbf{R}_k$ and $\mathbf{R}_l$ of the CC Hamiltonian matrix $\bar{\mathbf{H}}$ in the full manifold, with eigenvalues $E_k$ and $E_l$, and a positive-definite metric $M$ ($\mathbf{M}$ in matrix form). The two vectors span a two-dimensional subspace $V = \operatorname{span}\{\mathbf{R}_k, \mathbf{R}_l\}$ of the full manifold, for which we can construct an $M$-orthonormal basis $\{\mathbf{v}_1, \mathbf{v}_2\}$ ($\mathbf{v}_i^T \mathbf{M} \mathbf{v}_j = \delta_{ij}$). Since the subspace is spanned by eigenvectors of $\bar{\mathbf{H}}$, it is invariant under $\bar{\mathbf{H}}$. The representation of $\bar{\mathbf{H}}$ in this basis can generally be split into a symmetric and an anti-symmetric part as
\begin{equation}
    \bar{\mathbf{H}}^V =
    \begin{pmatrix}
        a & b \\
        c & d
    \end{pmatrix} =
    \begin{pmatrix}
        a & \frac{b+c}{2} \\
        \frac{b+c}{2} & d
    \end{pmatrix}
    + \zeta
    \begin{pmatrix}
        0 & 1 \\
        -1 & 0
    \end{pmatrix},
    \qquad \zeta = \frac{b-c}{2}.
    \label{eq:split}
\end{equation}
If $\zeta = 0$, the CC Hamiltonian matrix is symmetric in the basis $\{\mathbf{v}_1, \mathbf{v}_2\}$, and its
eigenvectors in the subspace are therefore $M$-orthogonal. Because $V$ is invariant under $\bar{\mathbf{H}}$, the eigenvectors of $\bar{\mathbf{H}}^V$ are also exact eigenvectors of $\bar{\mathbf{H}}$ in the full space. This means that, if the CC Hamiltonian matrix is symmetric in an $M$-orthonormal basis of the subspace spanned by two of its eigenvectors, these two eigenvectors are $M$-orthogonal.

For truncated CC methods, generally $\zeta \ne 0$ and the constrained states are not $M$-orthogonal. However, we can enforce $M$-orthogonality directly by introducing a correction to the full space Hamiltonian matrix which symmetrizes its representation in the subspace $V$. We call the corrected matrix the XO-CC Hamiltonian matrix
\begin{equation} \label{eq:DOCC_H}
    \tilde{\mathbf{H}} = \bar{\mathbf{H}} - \zeta \mathbf{C}.
\end{equation}
Here the XO-correction $\zeta\mathbf{C}$ is a full space matrix that, when projected onto the $M$-orthonormal basis $\{\mathbf{v}_1, \mathbf{v}_2\}$, transforms into the anti-symmetric matrix
\begin{equation}
    \mathbf{C}^V =
    \begin{pmatrix}
        0 & 1 \\
        -1 & 0
    \end{pmatrix}.
\end{equation}
This means that the XO-CC Hamiltonian matrix reduces to the symmetric part of the CC Hamiltonian matrix in the subspace $V$,
\begin{equation}
    \tilde{\mathbf{H}}^V = \operatorname{sym}(\bar{\mathbf{H}}^V).
\end{equation}
Accordingly, the eigenstates of $\tilde{\mathbf{H}}^V$ are $M$-orthogonal. By requiring $\mathbf{C}$ to map $V$ onto itself, we guarantee that $V$ is invariant under the XO-CC Hamiltonian. This ensures that the $M$-orthogonal XO-CC eigenvectors in the subspace are exact $M$-orthogonal eigenvectors also in the full space. These eigenvectors are then guaranteed to be real, have real eigenvalues and to have the correct crossing conditions for their intersections.

As the CC Hamiltonian matrix is real, but not symmetric, the constrained states may also be parallel or part of a complex pair. In these cases, the construction above is not well defined. Additionally, more than two states may become near-degenerate or cross at a single point. We therefore generalize the construction of the XO-CC Hamiltonian to treat all these cases. The spectral subspace of the CC Hamiltonian matrix associated with the $m$ constrained states (with eigenvalues $\{E_k\}$),
\begin{equation} \label{eq:spectral_subspace_V}
    V = \operatorname{null}\left[\prod_{k}(\bar{\mathbf{H}} - E_k\mathbf{1})\right],
\end{equation}
is an $m$-dimensional subspace invariant under $\bar{\mathbf{H}}$. It is uniquely defined as long as the eigenvalues $\{E_k\}$ differ from all other eigenvalues of $\bar{\mathbf{H}}$ \cite{LINALG}. We will now show that we can choose the constrained states such that $V$ is spanned by a real basis. Consider a complex pair $E_+$ and $E_-$. Then $E_+ + E_-$ and $E_+ E_-$ are real, and consequently the product 
\begin{equation}
    (\bar{\mathbf{H}} - E_+\mathbf{1})(\bar{\mathbf{H}} - E_-\mathbf{1}) = \bar{\mathbf{H}}^2 - (E_+ + E_-)\bar{\mathbf{H}} + E_+ E_-\mathbf{1}
\end{equation} 
is a real matrix. Thus, if $\{E_k\}$ is closed under complex conjugation, the product in Eq. \eqref{eq:spectral_subspace_V} is a real matrix (each pairwise product of the factors associated with a complex pair is real). The null space of a real matrix is always spanned by real vectors. Thus, as long as any complex eigenvector is constrained together with its partner, the spectral subspace of the Hamiltonian matrix associated with the constrained states is an invariant real subspace.

The spectral subspace is the natural real $m$-dimensional counterpart of the space spanned by the constrained states (which may be neither real nor $m$-dimensional). For real and linearly independent constrained states, $V$ is simply the span of their corresponding real eigenvectors. For any complex pair, $\mathbf{R_\pm} = \mathbf{R}^R \pm i \mathbf{R}^I$, $V$ is the span of their real and imaginary parts $\mathbf{R}^R$ and $\mathbf{R}^I$. This can be understood as $V$ containing the same information as the complex pair, but described by real vectors only. When two eigenstates become parallel at a defect (or exceptional point), they only span a one-dimensional space. Here, $V$ is the limit of the spaces for the linearly independent real or complex pairs as the exceptional point is approached from either side, and remains $m$-dimensional \cite{KATO}. Thus, the spectral subspace of the constrained states changes smoothly in the transition from linearly independent, to parallel and finally complex states (provided that $\bar{\mathbf{H}}$ itself is smooth, and the constrained states stay energetically separated from other states). We use this subspace to construct the XO-CC correction.

A real (but not $M$-orthonormal) basis $\{\mathbf{b}_i\}$ of $V$ is obtained as follows. Let $F$ be the dimensionality of the full manifold. We now construct the $F \times m$ matrix $\mathbf{B} = (\mathbf{b}_1 \, \dots \, \mathbf b_m)$ column-wise from $F \times 1$ dimensional submatrices $\mathbf{B}^k = (\mathbf{R}_k)$ for linearly independent states $\mathbf{R_k}$, $F \times 2$ dimensional $\mathbf{B}^{kk^*}=(\mathbf{R}_{k}^R, \mathbf{R}_{k}^I)$ for complex pairs $\mathbf{R}_k, \mathbf{R}_k^*$, and $F \times 2$ dimensional $\mathbf{B}^{kl}=(\mathbf{R}_{k}, \mathbf{G}_{l})$ for a pair of parallel states $\mathbf{R}_k =\mathbf{R}_l$ ($E_k = E_l$), where $(\bar{\mathbf{H}}-E_k\mathbf{1})\mathbf{G}_l = \mathbf{R}_k$. Correspondingly, we construct the $m \times m$ dimensional representation of $\bar{\mathbf{H}}$ in the basis $\mathbf{B}$, $\bar{\mathbf{H}}^B$, as the block-diagonal direct sum of 
\begin{equation} \label{eq:K}
    \mathbf{K}^k = (E_k), \qquad 
    \mathbf{K}^{kk^*}= 
    \begin{pmatrix}
        E_{k}^R & E_{k}^I \\
        -E_{k}^I & E_{k}^R \\
    \end{pmatrix}, \qquad 
    \mathbf{K}^{kl}= 
    \begin{pmatrix}
        E_{k} & 1 \\
        0 & E_{l} \\
    \end{pmatrix}
\end{equation}
for the same three cases, respectively. This choice fulfills $\bar{\mathbf{H}}\mathbf{B} = \mathbf{B}\bar{\mathbf{H}}^B$ for any number of constrained states \cite{KATO, LINALG}. The generalized eigenvector $\mathbf{G}_l$ is only determined up to the addition of a multiple of $\mathbf{R}_k$, but this does not change the span of the basis.

As the spectral subspace is $m$-dimensional, it has an $M$-orthonormal basis for any positive definite metric $M$. We construct such a basis by M-orthonormalizing $\mathbf{B}$. The metric enters through the $M$-overlap matrix (Gram matrix) $\Gamma_{ij} = \mathbf{b}_i^T\mathbf{M}\mathbf{b}_j$. For $M$ positive definite on $V$, and linearly independent $\{\mathbf{b}_i\}$, the $M$-overlap matrix is symmetric and positive definite, and as such invertible. Then the symmetric (Löwdin) orthonormalization \cite{lowdin1950non} $\mathbf{V} = \mathbf{B}\boldsymbol{\Gamma}^{-1/2}$ is well defined and gives an $M$-orthonormal basis of $V$, in which
\begin{equation}
    \bar{\mathbf{H}}^{V} = \boldsymbol{\Gamma}^{1/2}\bar{\mathbf{H}}^B\boldsymbol{\Gamma}^{-1/2}.
    \label{eq:H_V}
\end{equation}
We now expand the subspace Hamiltonian matrix into symmetric part and anti-symmetric parts as 
\begin{equation} \label{eq:H_V_multistate}
    \bar{\mathbf{H}}^V = \begin{pmatrix}
        d_1 & u_{12} & \dots \\
        l_{12} & d_2 & \dots \\
        \vdots & \vdots & \ddots \\
    \end{pmatrix} = 
    \begin{pmatrix}
        d_1 & s_{12} & \dots \\
        s_{12} & d_2 & \dots \\
        \vdots & \vdots & \ddots \\
    \end{pmatrix} +
    \begin{pmatrix}
        0 & \zeta'_{12} & \dots \\
        -\zeta'_{12} & 0 & \dots \\
        \vdots & \vdots & \ddots \\
    \end{pmatrix},
    \qquad 
    \begin{matrix}
        s_{ij} = \frac{u_{ij} + l_{ij}}{2} \\
        \zeta'_{ij} = u_{ij} - s_{ij},
    \end{matrix}
\end{equation}
which contains the anti-symmetric couplings $\zeta'_{ij}$ in the Löwdin basis. We construct the XO-CC Hamiltonian according to Eq. \eqref{eq:DOCC_H}. Our conditions on the correction matrix (subspace invariance and collapse to the anti-symmetric part of the CC Hamiltonian in the subspace) can be summarized as
\begin{equation} \label{eq:Ccond}
    \mathbf{C}\mathbf{V} = \mathbf{V} \bar{\mathbf{Z}}, \qquad \bar{\mathbf{Z}} \equiv \frac{\mathbf{Z}}{\zeta}, \qquad
    \mathbf{Z} \equiv \begin{pmatrix}
        0 & \zeta'_{12} & \dots \\
        -\zeta'_{12} & 0 & \dots \\
        \vdots & \vdots & \ddots \\
    \end{pmatrix},
\end{equation}
with the total anti-symmetric coupling defined as $\zeta \equiv \lVert\mathbf{Z}\rVert_F/\sqrt{2} = (\sum_{i<j} \zeta'^2_{ij})^{1/2}$. Only the product $\zeta\mathbf{C}$ enters the XO-CC Hamiltonian, where $\zeta$ sets the size of the correction and $\mathbf{C}$ its direction. For $\zeta = 0$ ($\mathbf{Z} = \mathbf{0}$), the correction vanishes, and we set $\mathbf{C} = \mathbf{0}$ directly. The representation of the XO-CC Hamiltonian matrix in the $M$-orthonormal basis becomes
\begin{equation}
    \tilde{\mathbf{H}}^V = \bar{\mathbf{H}}^V - \zeta\bar{\mathbf{Z}} =
\begin{pmatrix}
        d_1 & s_{12} & \dots \\
        s_{12} & d_2 & \dots \\
        \vdots & \vdots & \ddots \\
    \end{pmatrix},
\end{equation}
which is the symmetric part of Eq. \eqref{eq:H_V_multistate}. 

Equation \eqref{eq:Ccond} fixes $\mathbf{C}$ on $V$, but not on the remaining part of the full manifold. If we require that $\mathbf{C}$ maps every vector in the full manifold into $V$, so that the correction acts only within the subspace, $\mathbf{C}$ can be written as
\begin{equation}\label{eq:C}
    \mathbf{C} = \mathbf{V}\bar{\mathbf{Z}}\mathbf{Y},
\end{equation}
where the $m\times F$ matrix $\mathbf{Y}$ maps vectors in the full manifold into $V$. For Eq.~\eqref{eq:Ccond} to remain fulfilled, $\mathbf{Y}$ is required to be a left inverse of $\mathbf{V}$ ($\mathbf{Y}\mathbf{V} = \mathbf{1}_{m\times m}$). A natural choice is $\mathbf{Y} = (\mathbf{U}^T\mathbf{V})^{-1}\mathbf{U}^T$, where the columns of $\mathbf{U}$ span the left real spectral subspace 
\begin{equation}
    U = \operatorname{null}\left[\left(\prod_{k}(\bar{\mathbf{H}} - E_k\mathbf{1})\right)^T\right],
\end{equation}
of the constrained states. The matrix $\mathbf{U}$ can be constructed from the left EOM-CC states of the CC Hamiltonian in the same way as $\mathbf{B}$ is constructed from the right states. With this choice of $\mathbf{U}$, due to biorthogonality of the left and right EOM-CC states, $\operatorname{null} [\mathbf{Y}] = \operatorname{span}\{R_i\}, \, i \neq k$. Then, the right states other than the constrained ones are unchanged by the correction,
\begin{equation}
    \mathbf{C}\mathbf{R}_i = \mathbf{0} \implies \tilde{\mathbf{H}}\mathbf{R}_i = \bar{\mathbf{H}}\mathbf{R}_i, \qquad i \neq k.
\end{equation}
The left eigenvectors other than those of the constrained states are also unchanged, due to the same argument but with $\operatorname{null}[\mathbf{V}^T] = \operatorname{span}\{L_i\},\, i \neq k$,
\begin{equation}
    \mathbf{L}_i^T\mathbf{C} = \mathbf{0} \implies \mathbf{L}_i^T\tilde{\mathbf{H}} = \mathbf{L}_i^T\bar{\mathbf{H}}, \qquad i \neq k.
\end{equation}
Consequently, no states other than the constrained states are modified by the XO-CC correction. We also note that the total anti-symmetric coupling $\zeta$ is independent of the choice of $M$-orthonormal basis of $V$, as is the XO-CC Hamiltonian in the full manifold (and consequently all states and their properties). Finally, the XO correction $\zeta \mathbf{C}^V$ is the minimal change to $\bar{\mathbf{H}}^V$ (in the Frobenius norm) that $M$-orthogonalizes its eigenvectors, and preserves the average energy of the constrained states. These properties of XO-CC theory are derived in the Supplementary Materials.

The XO-CC right constrained states and their energies are determined by diagonalizing the real symmetric $m\times m$ matrix $\tilde{\mathbf{H}}^V$. The left modified states can similarly be determined, by diagonalizing the (generally nonsymmetric) $m\times m$ representation of $\tilde{\mathbf{H}}$ in the basis $\mathbf{U}$, $\tilde{\mathbf{H}}^U$, omitting the full-space EOM-CC eigenvalue problem. Expanding the $m$-dimensional left and right eigenvectors in their respective bases ($\mathbf{U}$ and $\mathbf{V}$) returns the full-space eigenvectors of the XO-CC Hamiltonian (up to biorthonormalization, see Supplementary Materials). The restoration of the correct crossing conditions for the constrained states can also be understood as the direct consequence of them being determined as the eigenvectors of the subspace Hamiltonian matrix $\tilde{\mathbf{H}}^V$, which is real and symmetric. This matrix then has the same crossing conditions for its eigenvectors as any real symmetric Hamiltonian.

The couplings $\zeta'_{ij}$ depend on the choice of $M$-orthonormal basis $\mathbf{V}$, and in the Löwdin basis they have no clear physical interpretation. A more useful basis is provided by the XO-CC constrained states, which themselves form an $M$-orthonormal basis of $V$. They are related to the Löwdin basis by an orthogonal transformation matrix $\mathbf{X}$. In this basis, the CC Hamiltonian takes the form
\begin{equation}
    \bar{\mathbf{H}}^X = \mathbf{X}^T\bar{\mathbf{H}}^V\mathbf{X} = \tilde{\mathbf{E}} + \bm{\zeta}, \qquad \tilde{\mathbf{E}} = \mathrm{diag}(\tilde{E}_1, \dots, \tilde{E}_m), \qquad \bm{\zeta} = \mathbf{X}^T\mathbf{Z}\mathbf{X},
    \label{eq:H_X}
\end{equation}
that is, its symmetric part is diagonal and contains the XO-CC energies $\tilde{E}_k$, and the states are coupled only through the anti-symmetric matrix $\bm{\zeta}$. We refer to the elements $\zeta_{ij}$ as the pairwise anti-symmetric couplings. They are the direct couplings between the constrained states which the XO-CC correction removes. As the XO-CC constrained states are only determined up to their signs, we report $|\zeta_{ij}|$. At degeneracies of XO-CC states, the eigenbasis is not unique, since any rotation among the degenerate states gives another eigenbasis. Such a rotation leaves the coupling between the degenerate states unchanged, but mixes their couplings to the remaining states. 

We will now consider the specific case of XO-CC theory for two constrained states. We can express the eigenvalues of the original CC Hamiltonian matrix in the spectral subspace using its matrix representation in the $M$-orthonormal basis $\mathbf{V}$ (see Eq. \eqref{eq:split}) as
\begin{equation}
    E_{k,l} = \frac{a+d}{2} \mp \sqrt{\delta^2 - \zeta^2}, \qquad
    \delta^2 = \left(\frac{a-d}{2}\right)^2 + \left(\frac{b+c}{2}\right)^2,
    \label{eq:Ekl}
\end{equation}
where $\pm\delta$ are the eigenvalues of the symmetric part relative to its mean. The eigenvalues are real for $\delta^2 > \zeta^2$ and complex for $\zeta^2 > \delta^2$. At $\delta^2 = \zeta^2$ the two eigenvalues coalesce and the two eigenvectors are parallel. On the other hand, the XO-CC energies of the constrained states are the eigenvalues of the symmetric part only,
\begin{equation}
    \tilde{E}_{k,l} = \frac{a+d}{2} \mp \delta.
    \label{eq:EklDO}
\end{equation}
The XO-CC states are degenerate only for $\delta = 0$, which requires both $a = d$ and $b + c = 0$. These are the same crossing conditions as in Hermitian theories, showing that the XO-CC intersections have the correct topology. Further, $\delta = 0$ implies that the CC constrained states either are degenerate or a complex pair, meaning the XO-CC intersection must lie on or within the intersection hypertube of the underlying CC method.

For more than two constrained states that are well separated, a perturbative expansion in the pairwise anti-symmetric couplings of Eq.~\eqref{eq:H_X} gives (see Supplementary Materials)
\begin{equation}
    \tilde{E}_k \approx E_k + \sum_{l \neq k} \frac{\zeta_{kl}^2}{\tilde{E}_k - \tilde{E}_l}.
    \label{eq:xo_second_order}
\end{equation}
Each pair of constrained states contributes separately to the correction, and the pairwise contributions to a pair of constrained states are equal and opposite. The anti-symmetric couplings are matrix elements of the anti-Hermitian part of the CC Hamiltonian, which arises from the truncation of the cluster operator and are generally not expected to grow with the energy difference of the states. Each state is therefore shifted away from another state it couples to by an amount inversely proportional to their energy difference. Then, the XO-CC modification of the constrained states decreases with their energetic separation, meaning that the correction is local in energy and acts strongest near intersections. Another way to interpret this result is that the anti-symmetric part of the Hamiltonian produces the largest numerical artifacts when the energy difference between the states is small. Thus, it is in the proximity of degeneracies that the XO-CC correction has the largest artifacts to correct.

While any metric that is positive definite on $V$ can be used, the use of the natural metric
\begin{equation} \label{eq:M_natural}
    M = \sum_{\phi}\exp{(T^\dagger)}\ket{\phi}\bra{\phi} \exp{(T)},
\end{equation}
where $\ket{\phi}$ is an orthonormal basis of the full manifold, guarantees the FCI limit (the untruncated coupled cluster Hamiltonian is symmetric in an $M$-orthonormal basis using this metric and hence $\zeta = 0$). On the other hand, the projected metric
\begin{equation} \label{eq:M_projected}
    M = \sum_{i,j}\exp{(T^\dagger)}\ket{b_i}\left[\bm{\mathcal{S}}^{-1}\right]_{ij}\bra{b_j}\exp{(T)}, \qquad \mathcal{S}_{ij} = \langle b_i\vert b_j\rangle,
\end{equation}
where $\ket{b_i}$ are the states corresponding to the basis vectors $\mathbf{b}_i$ of $V$, ensures the correct size-scaling properties. This is analogous to the metrics used in SCC theory \cite{SCCSD}.

Around ground state intersections XO-CC theory would suffer from the same bifurcations and negative excitation energies as the underlying CC method. On the other hand, GCC theory ensures continuous energies and avoids bifurcations in the solutions in these regions. However, GCC theory does not recover the correct crossing conditions for same-symmetry ground state intersections, and parallel states or complex energy pairs can be observed. By applying the XO-CC correction to the GCC states and energies, we can solve both problems in a unified theory, which we call XO-GCC theory. These XO-GCC methods describe ground state intersections without bifurcations, complex energies or parallel states, and recover the correct number of crossing conditions for the intersecting states.

We apply XO-CC theory to CC2, CCSD, GCC2 and GCCSD with a HF reference, so that the projection manifold consists of the singly and doubly excited singlet determinants. In practice, the XO-CC construction does not change the computational scaling of the underlying method, and XO-CCSD and XO-GCCSD scale as $N^6$, whereas XO-CC2 and XO-GCC2 scale as $N^5$, where $N$ is the number of molecular orbitals. Specifically, the additional steps of the XO-CC construction scale as $m^2N^4$ for $m$ constrained states, which is less than $N^5$ for $m < \sqrt{N}$, a condition fulfilled in practical applications.

\subsection*{Transition moments and oscillator strengths}
In both the GCC and the XO-CC/XO-GCC methods, the modifications to the underlying CC method are fully incorporated into the Hamiltonian matrix in the full manifold. Thus, any number of desired left and right EOM-CC states can be determined using this Hamiltonian. From the GCC, XO-CC and XO-GCC states we then determine electronic properties via a standard density matrix approach \cite{EOM-CC}. With the biorthonormalized left and right states, the one-electron transition density matrices are
\begin{equation}
    D^{kl}_{pq} = \langle L_k | E_{pq} | R_l \rangle
    = \sum_{\mu,\nu \geq 0} L_{k,\mu} \langle \mu | e^{-T} E_{pq} e^{T} | \nu \rangle R_{l,\nu},
    \label{eq:full_space_density}
\end{equation}
where the sums include the reference ($\mu,\nu = 0$) and $E_{pq}$ are the singlet excitation operators \cite{MEST}. The matrix elements $\langle \mu | e^{-T} E_{pq} e^{T} | \nu \rangle$ have the same form for all methods considered in this work, and are evaluated using the standard EOM-CCSD equations (though with different wave function parameters for the different methods). The transition moments in direction $\alpha$, $T^{\alpha}_{kl}$, and oscillator strengths $f_{0k}$, in the length gauge, are then determined as
\begin{equation}
    T^{\alpha}_{kl} = \sum_{pq} \mu^{\alpha}_{pq} D^{kl}_{pq},
    \qquad
    f_{0k} = \frac{2}{3}\,(\tilde{E}_k - \tilde{E}_0) \sum_{\alpha = x,y,z} T^{\alpha}_{0k} T^{\alpha}_{k0},
    \label{eq:osc_strength}
\end{equation}
where $\mu^{\alpha}_{pq}$ are the dipole integrals. This is the same form as in standard EOM-CC theory \cite{EOM-CC}.

\subsection*{Implementation details}\label{sec:impdet}

An XO-CC calculation proceeds in three steps. First, a reference calculation is performed. In this work we exclusively use a Hartree-Fock reference state, but a convex Hartree-Fock \cite{CVXHF} reference is also supported. Second, the CC or GCC ground and excited states are determined. Third, the XO-CC correction is constructed from the EOM-CC vectors of the underlying method, and the XO-CC right eigenvectors and eigenvalues are determined by the diagonalization of the subspace Hamiltonian matrix. The convergence of the XO-CC step is verified by evaluating the full-space residuals $\lVert\tilde{\mathbf{H}}\tilde{\mathbf{R}}_k - \tilde{E}_k\tilde{\mathbf{R}}_k\rVert$ of the XO-CC constrained states, which is a direct measure of the subspace invariance of $V$ and adds approximately the cost of one iteration in the Davidson solver for the underlying CC method. The $M$-orthogonality of the XO-CC constrained states does not need to be checked separately, as it holds by construction. The calculation is considered converged when all residuals are below the XO-CC threshold $r_\text{XO}$.

The states obtained from the underlying solvers may contain numerical errors from renormalization and projection steps, which can prevent the XO-CC residuals from reaching their threshold. Therefore, the left and right constrained states must be converged to sufficient accuracy in the underlying method. In our implementation we converge them to a residual threshold of $\min(r_\text{es}, 0.1\,r_\text{XO})$, where $r_\text{es}$ is the excited-state threshold in eT. This ensures that the subspace $V$ is spanned with the required accuracy needed to determine the XO-CC correction of the constrained states to the desired threshold.

The determination of the XO correction $\zeta \mathbf{C}$ requires $\bar{\mathbf{H}}^B$. For exact eigenvectors, $\bar{\mathbf{H}}^B$ is the block diagonal matrix built from the eigenvalues of the constrained states (see Eq. \eqref{eq:K}). In practice, the constrained states are converged only to a finite threshold, and the real and imaginary parts of complex pairs are normalized separately, meaning that the eigenvalue blocks do not represent $\bar{\mathbf{H}}$ in the actual stored basis $\mathbf{B}$. We therefore determine $\bar{\mathbf{H}}^B$ directly as the reduced representation
\begin{equation}
    \bar{\mathbf{H}}^B = (\mathbf{B}^T\mathbf{B})^{-1}\mathbf{B}^T\bar{\mathbf{H}}\mathbf{B},
    \label{eq:HB_lsq}
\end{equation}
which is independent of the normalization and phase conventions of the basis vectors. The products $\bar{\mathbf{H}}\mathbf{B}$ are retained from the final iteration of the convergence of the constrained states, such that no additional transformations with $\bar{\mathbf{H}}$ are required.

We have not implemented the generalized eigenvectors required for the treatment of exactly parallel eigenvectors. This means that, when moving arbitrarily close to the CC/GCC defect, the Gram matrix becomes numerically singular, hindering the accurate determination of the XO correction. In practice, we have not encountered points where our construction of $\mathbf{B}$ is insufficient to determine the XO correction to the desired threshold. If parallel constrained states become an issue, the generalized eigenvector approach could easily be added to the implementation.

For EOM-CC property calculations, any specified left and right XO-CC states are obtained from a Davidson solver \cite{davidson197514} on $\tilde{\mathbf{H}}$ and $\tilde{\mathbf{H}}^T$ respectively. For the right states, the closed-form XO-CC constrained states and remaining right eigenvectors of the underlying method are used as the start guess. For the left states, the left eigenvectors of the underlying method are used for all states. If no new states are introduced, the Davidson solvers then generally converge in a single iteration, as the subspace spanned by the start guesses contains the XO-CC states exactly.

In the original GCC implementation, the components of the $n$ projected states are removed from the cluster amplitudes by requiring $(\mathbf{l}^k)^T\mathbf{t} = 0$ for $k=1,\dots,n$. However, the doubles parts of the cluster and EOM-CC vectors are stored in different conventions,
\begin{equation}
    \ket{t_\text{doubles}} = \sum_{ai\leq bj} \frac{1}{1+\delta_{ai,bj}} t_{ij}^{ab} \ket{\mu_{ij}^{ab}},
    \qquad
    \bra{L^k_\text{doubles}} = \sum_{ai\leq bj} l_{ij}^{ab,k} \bra{\mu_{ij}^{ab}},
\end{equation}
such that $(\mathbf{l}^k)^T\mathbf{t} = \braket{L^k}{t} + \frac{1}{2}\sum_{ai} l_{ii}^{aa,k}\, t_{ii}^{aa}$, i.e. the diagonal doubles are counted twice. The extra term is not invariant to occupied-occupied and virtual-virtual orbital rotations, and the GCC energies obtained with this constraint therefore depend on the choice of orbitals within these spaces \cite{stoll2026cvxscc}. In this work, we instead impose $\braket{L^k}{t} = 0$, which is invariant to such rotations. For GCCSD, we verified that the energies obtained with this projection are invariant to occupied-occupied and virtual-virtual rotations.

%%%%%%%%%%%%%%%% MAIN TEXT FIGURES %%%%%%%%%%%%%%%

%%%%%%%%%%%%%%%% MAIN TEXT TABLES %%%%%%%%%%%%%%%

%%%%%%%%%%%%%%%% REFERENCES %%%%%%%%%%%%%%%

\clearpage % Clear all remaining figures and tables then start a new page

% The list of references goes after the main text and before the acknowledgements
% When preparing an initial submission, we recommend you use BibTeX, like this:
%
\bibliography{XO_CC} % for a file named science_template.bib
\bibliographystyle{sciencemag}

% After the paper has completed peer review and been revised ready for acceptance,
% you should comment out the lines above and copy-paste the contents of your .bbl
% file here instead. This will help ensure that our conversion software works correctly.
% Remember to re-run BibTeX first - check the timestamp!
%
% Example of the first three entries copy-pasted from science_template.bbl:
%
%\begin{thebibliography}{1}
%
%\bibitem{example}
%A.~N. {Author}, An example reference. \emph{Journal of Improbable Research}
%  \textbf{1}, 67 (2020).
%
%\bibitem{example2}
%F.~M. {Surname}, S.~{Author}, A second example. \emph{Interesting Research
%  Letters} \textbf{32}, 897 (2019).
%
%\bibitem{example_preprint}
%P.~{One}, P.~{Two}, P.~{Three}, {An unpublished preprint}. \emph{preprint}
%  (2021), arXiv:2101.12345.
%
%\end{thebibliography}

%%%%%%%%%%%%%%%% ACKNOWLEDGEMENTS %%%%%%%%%%%%%%%

\section*{Acknowledgments}
We would like to thank Eirik F. Kjønstad for his contributions to metric-related routines which were reused in our implementation.

\paragraph*{Funding:}
This work was supported by the European Research Council (ERC) under the European Union's Horizon 2020 Research and Innovation Program (grant agreement No. 101020016).
\paragraph*{Author contributions:} 
L.S. and H.K. conceived the project. L.S. and F.R. developed the implementation in eT. L.S. performed all the calculations. L.S. wrote the first draft of the paper, and all authors reviewed and approved the final version. H.K. supervised the project and secured funding.

\paragraph*{Competing interests:}
There are no competing interests to declare.

\paragraph*{Data and materials availability:}
Calculation results are available at https://zenodo.org/records/24134777

\paragraph*{Use of LLMs}
The large language model Claude Opus 5.5 (Anthropic) was used as an aid in this work: to develop the Python driver for the finite-difference MECI optimizations, the in-plane CI determinations, and scripts for data analysis and plotting, to draft parts of the Supplementary Materials text, to check derivations and language in the manuscript, and to assist in the final implementation in eT. All AI-assisted code and text were reviewed, edited and verified by the authors, who take full responsibility for the content.

%%%%%%%%%%%%%%%% SUPPLEMENT LIST %%%%%%%%%%%%%%%

% List the contents of your Supplementary Materials, including the numbers of any
% supplementary figures, tables, external data files etc. and any references that are
% cited only in the supplement. In this example, refs. 7-8 are cited only in the supplement.
% Fill out your numbers accordingly and delete any lines that aren't applicable.
\subsection*{Supplementary materials}
Supplementary Text\\
Figures S1 to S2\\
Tables S1 to S2\\
% References \textit{(7-\arabic{enumiv})}\\ % automatically fills out the last reference number
% (filling out the other numbers automatically is possible but fiddly and liable to break)
% Movie S$_1$\\
% Data S$_1$

%%%%%%%%%%%%%%%% END OF MAIN TEXT %%%%%%%%%%%%%%%

\newpage

%%%%%%%%%%%%%%%% START OF SUPPLEMENT %%%%%%%%%%%%%%%

% Figures, tables, equations and pages in the supplement are numbered S$_1$, S2 etc.
\renewcommand{\thefigure}{S\arabic{figure}}
\renewcommand{\thetable}{S\arabic{table}}
\renewcommand{\theequation}{S\arabic{equation}}
\renewcommand{\thepage}{S\arabic{page}}
\setcounter{figure}{0}
\setcounter{table}{0}
\setcounter{equation}{0}
\setcounter{page}{1} % not 0 as \newpage already started a supplementary page
% References continue the numbering from the main text.

%%%%%%%%%%%%%%%% SUPPLEMENT TITLE PAGE %%%%%%%%%%%%%%%

\begin{center}
\section*{Supplementary Materials for\\ \scititle}

% Author list for the supplement
% Indicate the corresponding authors, but do NOT include institutions here
% It would be nice if the template auto-generated this, but doing so is complicated...
	% You can write out first names or use initials - either way is acceptable, but be consistent
	Leo Stoll$^{1}$,
	Federico Rossi$^{1}$ and
	Henrik Koch$^{1\ast}$ \\
	% Additional lines of authors should be inserted using the \and command (not \\)
	% Institution list, in a slightly smaller font
	\small$^{1}$Department of Chemistry, Norwegian University of Science and Technology, Trondheim 7491, Norway. \\
	% Identify at least one corresponding author, with contact email address
	\small$^\ast$Corresponding author. Email: henrik.koch@ntnu.no \\
\end{center}

% Fill out the numbers for each type of supplementary material,
% and delete any lines that aren't applicable.
% These are just example numbers that don't match the rest of this template.
\subsection*{This PDF file includes:}
Supplementary Text\\
Figures S1 to S2\\
Tables S1 to S2\\
% Captions for Movies S$_1$ to S2\\
% Captions for Data S$_1$ to S2

% \subsubsection*{Other Supplementary Materials for this manuscript:}
% Movies S$_1$ to S2\\
% Data S$_1$ to S2

\newpage

%%%%%%%%%%%%%%%% SUPPLEMENTARY TEXT %%%%%%%%%%%%%%%

\subsection*{Supplementary text}

\subsubsection*{Additional derivations on XO-CC theory}

In this section we show that, for an arbitrary number $m$ of constrained states, (i) the total anti-symmetric coupling $\zeta$ and (ii) the XO-CC Hamiltonian matrix $\tilde{\mathbf{H}}$ do not depend on the choice of the $M$-orthonormal basis $\mathbf{V}$, that (iii) the XO-CC correction preserves the average energy of the constrained states, that (iv) for well-separated constrained states, the effect of the correction on their energies is inversely proportional to their energy difference, and finally that (v) the correction $\zeta \mathbf{C}$ can be interpreted as the minimal correction to $\bar{\mathbf{H}}$ that makes the constrained states $M$-orthogonal.

\paragraph{(i) The total anti-symmetric coupling is independent of the basis.}
Any two $M$-orthonormal bases $\mathbf{V}$ and $\mathbf{V}'$ of $V$ are related by an invertible change of basis transformation $\mathbf{Q}$. The $M$-orthonormality implies
\begin{equation}
    \mathbf{1} = \mathbf{V}'^T\mathbf{M}\mathbf{V}' = \mathbf{Q}^T\left(\mathbf{V}^T\mathbf{M}\mathbf{V}\right)\mathbf{Q} = \mathbf{Q}^T\mathbf{Q},
\end{equation}
so $\mathbf{Q}$ is an orthogonal matrix, $\mathbf{Q}^T = \mathbf{Q}^{-1}$. In other words, any two $M$-orthonormal bases of $V$ are related by rotations and reflections. Inserting $\mathbf{V} = \mathbf{V}'\mathbf{Q}^T$ gives
\begin{equation}
    \bar{\mathbf{H}}\mathbf{V}' = \mathbf{V}'\bar{\mathbf{H}}^{V'}, \qquad
    \bar{\mathbf{H}}^{V'} = \mathbf{Q}^T\bar{\mathbf{H}}^V\mathbf{Q} = \mathbf{Q}^T\mathbf{S}\mathbf{Q} + \mathbf{Q}^T\mathbf{Z}\mathbf{Q},
\end{equation}
where $\mathbf{S}$ and $\mathbf{Z}$ denote the symmetric and anti-symmetric components of $\bar{\mathbf{H}}^V$. The matrix $\mathbf{Q}^T\mathbf{S}\mathbf{Q}$ is again symmetric and $\mathbf{Q}^T\mathbf{Z}\mathbf{Q}$ is again anti-symmetric. As the splitting into a symmetric and an anti-symmetric part is unique, the two parts in the new basis are
\begin{equation}
    \mathbf{S}' = \mathbf{Q}^T\mathbf{S}\mathbf{Q}, \qquad \mathbf{Z}' = \mathbf{Q}^T\mathbf{Z}\mathbf{Q}.
    \label{eq:SI_transform}
\end{equation}
The change of basis thus mixes the elements of the anti-symmetric part. For example, a rotation in the plane of the basis vectors $\mathbf{v}_1$ and $\mathbf{v}_2$ leaves the $(1,2)$ element unchanged, but mixes the $(1,3)$ and $(2,3)$ elements. This is why the couplings $\zeta'_{ij}$ of the Löwdin basis have no physical meaning on their own, and why we report the couplings $\zeta_{ij}$ in the eigenbasis of $\mathbf{S}$ (Eq.~\eqref{eq:H_X}).

Using Eq.~\eqref{eq:SI_transform}, $\mathbf{Q}\mathbf{Q}^T = \mathbf{1}$ and the cyclic property of the trace,
\begin{equation}
    \lVert\mathbf{Z}'\rVert_F^2 = \mathrm{tr}\left(\mathbf{Z}'^T\mathbf{Z}'\right)
    = \mathrm{tr}\left(\mathbf{Q}^T\mathbf{Z}^T\mathbf{Q}\mathbf{Q}^T\mathbf{Z}\mathbf{Q}\right)
    = \mathrm{tr}\left(\mathbf{Z}^T\mathbf{Z}\mathbf{Q}\mathbf{Q}^T\right)
    = \lVert\mathbf{Z}\rVert_F^2 .
\end{equation}
Hence $\lVert\mathbf{Z}'\rVert_F = \lVert\mathbf{Z}\rVert_F$, i.e. the total anti-symmetric coupling $\zeta$ is the same in every $M$-orthonormal basis of $V$.

\paragraph*{(ii) The XO-CC Hamiltonian is independent of the basis.}
We first note that $\mathbf{Y}$ does not depend on the choice of basis for $U$. Replacing $\mathbf{U}$ by $\mathbf{U}\mathbf{P}$ with any invertible $m\times m$ matrix $\mathbf{P}$ gives
\begin{equation}
    \left(\mathbf{P}^T\mathbf{U}^T\mathbf{V}\right)^{-1}\mathbf{P}^T\mathbf{U}^T
    = (\mathbf{U}^T\mathbf{V})^{-1}\left(\mathbf{P}^T\right)^{-1}\mathbf{P}^T\mathbf{U}^T
    = \mathbf{Y}.
\end{equation}
Under the change of basis $\mathbf{V}' = \mathbf{V}\mathbf{Q}$, the left inverse transforms as
\begin{equation}
    \mathbf{Y}' = \left(\mathbf{U}^T\mathbf{V}\mathbf{Q}\right)^{-1}\mathbf{U}^T = \mathbf{Q}^T(\mathbf{U}^T\mathbf{V})^{-1}\mathbf{U}^T = \mathbf{Q}^T\mathbf{Y}.
\end{equation}
Combining this with Eq.~\eqref{eq:SI_transform}, all factors of $\mathbf{Q}$ cancel in the correction,
\begin{equation}
    \zeta\mathbf{C}' = \mathbf{V}'\mathbf{Z}'\mathbf{Y}'
    = \mathbf{V}\mathbf{Q}\,\mathbf{Q}^T\mathbf{Z}\mathbf{Q}\,\mathbf{Q}^T\mathbf{Y}
    = \mathbf{V}\mathbf{Z}\mathbf{Y}
    = \zeta\mathbf{C}.
\end{equation}
The XO-CC Hamiltonian matrix $\tilde{\mathbf{H}} = \bar{\mathbf{H}} - \zeta\mathbf{C}$ is therefore uniquely determined by the choice of constrained states and the metric $M$. Consequently, all XO-CC energies, left and right states, and properties computed from them are independent of how the bases of $U$ and $V$ are chosen. This holds even though the individual matrices $\mathbf{V}$, $\mathbf{Y}$ and $\mathbf{Z}$ do depend on that choice.

\paragraph*{(iii) The average energy of the constrained states is preserved.}
The sum of the eigenvalues of a matrix equals its trace. The $m$ constrained CC energies $E_k$ are the eigenvalues of $\mathbf{K}$, and by Eq.~\eqref{eq:H_V} also of $\bar{\mathbf{H}}^V$, which is related to $\mathbf{K}$ by a similarity transformation with the $M$-overlap matrix. The XO-CC energies $\tilde{E}_k$ are the eigenvalues of $\mathbf{S}$. Since the anti-symmetric matrix $\mathbf{Z}$ is traceless,
\begin{equation}
    \sum_{k}\tilde{E}_k = \mathrm{tr}\,\mathbf{S} = \mathrm{tr}\left(\bar{\mathbf{H}}^V - \mathbf{Z}\right) = \mathrm{tr}\,\bar{\mathbf{H}}^V = \mathrm{tr}\,\mathbf{K} = \sum_{k}E_k,
\end{equation}
where the sums run over the $m$ constrained states. The average energy of the constrained states is therefore unchanged by the XO-CC correction.

\paragraph*{(iv) For well-separated constrained states, the effect of the correction on their energies is inversely proportional to their energy difference.}
In the eigenbasis of $\mathbf{S}$, the CC Hamiltonian of the constrained states is the XO-CC Hamiltonian $\tilde{\mathbf{E}}$ plus the anti-symmetric coupling $\bm{\zeta}$ that the correction removes (Eq.~\eqref{eq:H_X}). We can therefore view the CC Hamiltonian $\bar{\mathbf{H}}^X = \tilde{\mathbf{E}} + \bm{\zeta}$ as the diagonal matrix $\tilde{\mathbf{E}}$ of XO-CC energies, perturbed by the small anti-symmetric couplings $\bm{\zeta}$. Switching on the couplings turns the XO-CC energies $\tilde{E}_k$ into the CC energies $E_k$, and perturbation theory tells us how large this change is. It helps to compare with the familiar symmetric case. In second order, a coupling between two states shifts their energies through a transition from state $k$ to state $l$ and back, and the shift is therefore proportional to the product of the two off-diagonal elements, $\zeta_{kl}\zeta_{lk}$, divided by the energy difference (this holds also for nonsymmetric matrices). For a symmetric coupling, this product is positive, and the levels are pushed apart. For the anti-symmetric coupling, $\zeta_{lk} = -\zeta_{kl}$, so the product is $-\zeta_{kl}^2$. The shift has the same size as for a symmetric coupling, but the opposite sign: the anti-symmetric coupling pulls the two levels together. This is why CC pairs can coalesce and become complex. There is no first-order shift, since $\bm{\zeta}$ has no diagonal elements. Adding the contributions of all pairs gives
\begin{equation}
    E_k = \tilde{E}_k - \sum_{l\neq k}\frac{\zeta_{kl}^2}{\tilde{E}_k - \tilde{E}_l} + \mathcal{O}(\zeta^3),
    \label{eq:SI_PT2}
\end{equation}
which is Eq.~\eqref{eq:xo_second_order} of the main text. Read in reverse, the XO-CC correction removes the couplings and pushes each pair of states apart by the same amount.

The elements of $\bm{\zeta}$ arise from the truncation of the cluster operator, and are not expected to grow with the energy difference of the states. This establishes the locality of the correction. Constrained states that are well separated in energy change each other's energies only at second order in $\zeta_{kl}$, and inversely proportional to their energy difference. 
 
 The expansion requires each coupling to be small compared to the energy difference it is divided by, $|\zeta_{kl}| \ll |\tilde{E}_k - \tilde{E}_l|$. Near an intersection, the XO-CC energy difference goes to zero while the coupling stays finite, so the second-order term in Eq.~\eqref{eq:SI_PT2} becomes large and the expansion breaks down for the intersecting pair. Such a pair is instead treated exactly by Eqs.~\eqref{eq:Ekl} and \eqref{eq:EklDO}. The couplings to the remaining constrained states still contribute only second-order shifts of the form of Eq.~\eqref{eq:SI_PT2}, so that the XO-CC correction to the intersecting pair is essentially that of the two-state case.

 \paragraph*{(v) The correction $\zeta \mathbf{C}$ can be interpreted as the minimal correction to $\bar{\mathbf{H}}$ that makes the constrained states $M$-orthogonal.}
 Within $V$, the correction changes $\bar{\mathbf{H}}^V = \mathbf{S} + \mathbf{Z}$ by $-\zeta\mathbf{C}^V = -\zeta\bar{\mathbf{Z}} = -\mathbf{Z}$, that is, it removes exactly the anti-symmetric part of $\bar{\mathbf{H}}^V$. Symmetric and anti-symmetric matrices are orthogonal in the Frobenius inner product, so that for any symmetric $m\times m$ matrix $\mathbf{S}'$
\begin{equation}
    \lVert \bar{\mathbf{H}}^V - \mathbf{S}' \rVert_F^2 = \lVert \mathbf{S} - \mathbf{S}' \rVert_F^2 + \lVert \mathbf{Z} \rVert_F^2 = \lVert \mathbf{S} - \mathbf{S}' \rVert_F^2 + 2\zeta^2,
\end{equation}
where $\mathbf{S}$ is the symmetric part of $\bar{\mathbf{H}}^V$. The lowest norm is obtained with $\mathbf{S}' = \mathbf{S}$, which is the case for the XO-CC Hamiltonian ($\tilde{\mathbf{H}}^V = \mathbf{S}$). In other words, the XO correction $-\zeta\mathbf{C}^V = -\mathbf{Z}$ is the smallest change (in the Frobenius norm) of $\bar{\mathbf{H}}^V$ that makes it symmetric, and thereby the constrained states $M$-orthogonal. Since the Frobenius norm and the symmetry of a matrix are preserved under orthogonal transformations, this statement does not depend on the choice of $M$-orthonormal basis. In this sense, $\zeta\mathbf{C}$ is a minimal correction: it is the smallest possible change that symmetrizes the Hamiltonian in the subspace $V$, and in this way achieves $M$-orthogonality of the constrained states. Outside the subspace, the correction is minimal in a different sense: it leaves all other states and energies unchanged.

\newpage

\subsubsection*{Uracil MECIs}

\textit{MECI optimization}\\
The cc-pVDZ minimum energy conical intersections (MECIs) of uracil were located with the tube algorithm of Angelico \textit{et al.}~\cite{angelico2025determining}. In this algorithm the energy of the upper state is minimized on the surface on which the two intersecting states are separated by a small, constant energy difference $\varepsilon$; on this surface, this is equivalent to minimizing their mean energy $\bar{E}$. The resulting $\varepsilon$-MECI approaches the MECI as $\varepsilon \to 0$. The algorithm needs only the energies of the two states, the gradient of their energy average $\bar{E}$ and the gradient of their energy difference $\Delta E$. Derivative couplings are not required, so the branching plane does not have to be constructed during the optimization. The S$_1$/S$_2$ MECIs were determined with XO-CC2 and XO-CCSD, and the $^6S_5$ S$_0$/S$_1$ MECIs with XO-GCC2 and XO-GCCSD. The GCC methods used one projected state, and the two intersecting states were the constrained states of the XO step. The starting geometries were the SCCSD S$_1$/S$_2$ gradient projection MECI and the CCSD $^6S_5$ S$_0$/S$_1$ $\varepsilon$-MECI ($\varepsilon = 0.0025~E_h$) of Ref.~\cite{angelico2025determining}. All electronic structure calculations were performed with eT~\cite{eT}. The optimization code was developed with the assistance of the large language model Claude Opus 5.5 (Anthropic), reviewed by the authors, and validated on analytical model surfaces with known MECIs before production use.

Analytical XO-CC gradients are not yet available, so all gradients were obtained by central finite differences of the energies. The displacements were taken along the 30 ($3N-6$) orthonormal Cartesian directions that remain after rigid translations and rotations are projected out. Every gradient therefore requires a stencil of 60 displaced geometries plus the reference geometry, 61 single-point calculations in total. Two quantities were differenced: $\bar{E}$ and $(\Delta E)^2$. The unit normal $\hat{\mathbf{n}}$ of the tube surface $\Delta E = \varepsilon$, i.e.\ the direction of the gradient of $\Delta E$, and the projected gradient follow from these two gradients. The finite-difference step was reduced together with $\varepsilon$, see Table \ref{tab:sup_Uracil_MECI_stages}.

In each iteration, a Newton step on $\Delta E$ was taken along $\hat{\mathbf{n}}$ and a quasi-Newton step on $\bar{E}$ (BFGS update of the Hessian). The step component that rotates $\hat{\mathbf{n}}$ was limited to $0.2$ times the tube radius, and a second-order correction along $\hat{\mathbf{n}}$ accounted for the curvature of $\Delta E$. The step within the seam was limited to $0.5\,(\varepsilon/E_h)^{1/2}$~bohr. From the second stage of each optimization on, this limit was doubled when the previous step had been limited, $\Delta E$ was within 1\% of $\varepsilon$, and the decrease of $\bar{E}$ was at least half of the predicted decrease; otherwise it was reset.

As in Ref.~\cite{angelico2025determining}, $\varepsilon$ was reduced stepwise, and each stage was started from the converged $\varepsilon$-MECI of the previous stage (Table~\ref{tab:sup_Uracil_MECI_stages}). For XO-CCSD the first stage was omitted, because XO-CCSD is closely related to SCCSD, and its S$_1$/S$_2$ MECI is therefore expected close to the SCCSD starting geometry (the XO-CCSD energy difference at the starting geometry is 0.077~m$E_h$). A stage was considered converged when the maximum and root-mean-square components of the projected gradient were below the thresholds in Table~\ref{tab:sup_Uracil_MECI_stages} and $\Delta E$ was within 5\% of $\varepsilon$. For $\varepsilon \le 10^{-4}~E_h$, the change of the upper-state energy in the last iteration and its predicted remaining decrease also had to be below the energy tolerance in Table~\ref{tab:sup_Uracil_MECI_stages}.

\begin{table}[h!]
\centering
\caption{\textbf{Stepwise $\varepsilon$ schedule of the uracil MECI optimizations.} $\delta$ is the finite-difference step. The gradient thresholds refer to the maximum (max) and root-mean-square (rms) components of the projected gradient in $E_h$/bohr; the energy tolerance applies to the change of the upper-state energy and to its predicted remaining decrease. The last columns give the number of iterations per stage.}
\label{tab:sup_Uracil_MECI_stages}
\resizebox{\textwidth}{!}{%
\begin{tabular}{c c c c c c c c}
\hline
 & & & & \multicolumn{4}{c}{Iterations} \\
$\varepsilon$ ($E_h$) & $\delta$ (\AA) & max / rms ($E_h$/bohr) & Energy tol.\ ($E_h$) & XO-CC2 & XO-CCSD & XO-GCC2 & XO-GCCSD \\
\hline
$10^{-2}$ & 0.003 & $3\cdot10^{-3}$ / $1.2\cdot10^{-3}$ & -- & 8 & -- & 14 & 12 \\
$10^{-3}$ & 0.002 & $3\cdot10^{-4}$ / $1.2\cdot10^{-4}$ & -- & 72 & 12 & 33 & 13 \\
$10^{-4}$ & 0.00063 & $10^{-4}$ / $5\cdot10^{-5}$ & $10^{-6}$ & 32 & 11 & 14 & 12 \\
$3\cdot10^{-5}$ & 0.00035 & $10^{-4}$ / $5\cdot10^{-5}$ & $5\cdot10^{-7}$ & 20 & -- & -- & -- \\
\hline
\multicolumn{4}{l}{Total} & 132 & 23 & 61 & 37 \\
\hline
\end{tabular}}
\end{table}

\noindent{\textit{MECI energies and the $\varepsilon \to 0$ limit}}\\
Close to the intersection, in the branching plane the two PESs form a tilted double cone. With displacements $x$ and $y$ (in bohr) along the branching-plane vectors $\mathbf{g}$ and $\mathbf{h}$ (see below),
\begin{equation}
  E_\pm = E_\mathrm{X} + s_g x + s_h y \pm \sqrt{|\mathbf{g}|^2x^2 + |\mathbf{h}|^2y^2},
\end{equation}
where $E_\mathrm{X}$ is the MECI energy and $s_g$ and $s_h$ are the in-plane components of the gradient of $\bar{E}$ (the tilt of the cone). Minimizing $\bar{E}$ on the ellipse $\Delta E = \varepsilon$ gives
\begin{equation}
  \bar{E}(\varepsilon) = E_\mathrm{X} - \kappa\,\varepsilon + \mathcal{O}(\varepsilon^2),
  \qquad
  \kappa = \frac{1}{2}\sqrt{\left(\frac{s_g}{|\mathbf{g}|}\right)^2 + \left(\frac{s_h}{|\mathbf{h}|}\right)^2},
\end{equation}
where the $\varepsilon^2$ term arises from the curvature of the seam and the relaxation of the $\varepsilon$-MECI along it. The mean energy at an $\varepsilon$-MECI therefore lies below the MECI energy by approximately $\kappa\varepsilon$, which amounts to 7-26~$\mu E_h$ at $\varepsilon = 10^{-4}~E_h$ for the uracil MECIs (Table~\ref{tab:sup_Uracil_MECI_energies}). The MECI energies and geometries were therefore obtained by linear extrapolation in $\varepsilon$ from the last two stages, $E_\mathrm{X} \approx \bar{E}(\varepsilon_2) + [\bar{E}(\varepsilon_2) - \bar{E}(\varepsilon_1)]\,\varepsilon_2/(\varepsilon_1 - \varepsilon_2)$, and analogously for the Cartesian coordinates. This extrapolation is a numerical step of the optimization; it requires only that the intersection is conical, which is the case for XO-CC.

The extrapolation was validated with numerically degenerate points located in the branching plane (see below) of each extrapolated MECI. Every point on the seam lies at or above the MECI, so the mean energy at such a point is an upper bound to the MECI energy. For XO-CCSD, XO-GCC2 and XO-GCCSD, the extrapolation from $\varepsilon = 10^{-3}$ and $10^{-4}~E_h$ gives energies 0.8, 0.7 and 0.9~$\mu E_h$ below these upper bounds. For XO-CC2, the same extrapolation gives $-413.5572826431~E_h$, which lies 1.2~$\mu E_h$ above the corresponding seam point and thus overestimates the MECI energy. The XO-CC2 stage with $\varepsilon = 10^{-3}~E_h$ is not yet in the linear regime: its geometry differs from the $\varepsilon = 10^{-4}~E_h$ geometry by 0.096~\AA, compared with about 0.01~\AA\ for the other methods. An additional stage with $\varepsilon = 3\cdot10^{-5}~E_h$ was therefore performed, and the XO-CC2 MECI was extrapolated from $\varepsilon = 10^{-4}$ and $3\cdot10^{-5}~E_h$. This value lies 0.07~$\mu E_h$ below the seam point. We therefore consider the reported MECI energies converged to about $1~\mu E_h$. The stage energies, the extrapolated MECI energies and the seam-point energies are collected in Table~\ref{tab:sup_Uracil_MECI_energies}. All MECI geometries are available in the data repository.\\

\noindent{\textit{Branching plane and seam points}}\\
XO-CC currently has no analytical nuclear gradients or derivative couplings, so the branching plane of each MECI was determined from energies alone, around the extrapolated MECI geometry $\mathbf{R}_0$. To first order in the displacements, the energy difference between the two intersecting XO states is
\begin{equation}
  \Delta E(\mathbf{q}) = \left| \mathbf{w}(\mathbf{q}) \right|, \qquad
  \mathbf{w}(\mathbf{q}) = \mathbf{w}_0 + \mathbf{G}^{T}\mathbf{q},
\end{equation}
where $\mathbf{q}$ contains the displacements from $\mathbf{R}_0$ along 30 orthonormal internal (translation- and rotation-free) Cartesian directions, the two-component vector $\mathbf{w}_0$ gives the gap at $\mathbf{R}_0$ ($\Delta E(\mathbf{0}) = |\mathbf{w}_0|$), and the two columns of the $30\times2$ matrix $\mathbf{G}$ span the branching plane. Both components of $\mathbf{w}$ vanish on the seam, which to first order is determined as $\mathbf{q}_X = -\mathbf{G}(\mathbf{G}^T\mathbf{G})^{-1}\mathbf{w}_0$.

To determine $\mathbf{w}_0$ and $\mathbf{G}$, we performed calculations at $\mathbf{R}_0$ and at displacements of $\pm\delta$, with $\delta = 2\cdot10^{-4}$~\AA, along each of the 30 directions. Because $\mathbf{w}_0 \neq \mathbf{0}$, these central differences determine the component of each row of $\mathbf{G}$ along $\mathbf{w}_0$, but only the magnitude of its component perpendicular to $\mathbf{w}_0$. The relative signs of the perpendicular components were determined from 29 additional calculations with simultaneous displacements along a reference direction, chosen as the one with the largest perpendicular component, and each of the other directions. The remaining overall sign corresponds to a reflection within the branching plane and does not affect it. The model was then fitted to the gaps at all 90 geometries, with a root-mean-square deviation below $5\cdot10^{-5}$~meV. A singular value decomposition of $\mathbf{G}$ brings the model into the standard form
\begin{equation}
  \Delta E = 2\sqrt{(\mathbf{g}\cdot\Delta\mathbf{R})^2 + (\mathbf{h}\cdot\Delta\mathbf{R})^2},
\end{equation} 
where $\Delta\mathbf{R} = \mathbf{q} - \mathbf{q}_X$ is the displacement from the intersection, and $\mathbf{g}$ and $\mathbf{h}$ are the two left singular vectors of $\mathbf{G}$, scaled by half the corresponding singular values. Energies alone determine the cone, but not the individual gradient-difference and coupling vectors, which depend on the choice of the two degenerate states and are in general not orthogonal. $\mathbf{g}$ and $\mathbf{h}$ obtained here are the orthogonal pair to which they can always be rotated; this pair is unique up to the signs of the vectors, and we denote the longer one by $\mathbf{g}$. Note that $\mathbf{h}$ is not obtained from a derivative coupling; it is the second direction of the branching plane, determined from energies alone like $\mathbf{g}$. The vectors are provided in the data repository.

\begin{table}[tbp!]
\centering
\caption{\textbf{Mean energies $\bar{E}$ ($E_h$) of the uracil $\varepsilon$-MECIs and extrapolated MECIs (cc-pVDZ).} The MECI energy is the linear $\varepsilon \to 0$ extrapolation from the last two stages ($\varepsilon = 10^{-3}$ and $10^{-4}~E_h$; for XO-CC2 $10^{-4}$ and $3\cdot10^{-5}~E_h$). The seam point is the exactly degenerate point in the branching plane of the extrapolated MECI geometry, and its mean energy is an upper bound to the exact MECI energy. For the extrapolated MECI and the seam point, the energy gap between the intersecting states is given in $E_h$. $\Delta$ is the seam-point energy relative to the MECI energy.}
\label{tab:sup_Uracil_MECI_energies}
\resizebox{\textwidth}{!}{%
\begin{tabular}{l c c c c}
\hline
 & \multicolumn{2}{c}{S$_1$/S$_2$} & \multicolumn{2}{c}{$^6S_5$ S$_0$/S$_1$} \\
 & XO-CC2 & XO-CCSD & XO-GCC2 & XO-GCCSD \\
\hline
$\bar{E}(\varepsilon = 10^{-2})$ & $-413.5588179823$ & -- & $-413.5947089806$ & $-413.5951528566$ \\
$\bar{E}(\varepsilon = 10^{-3})$ & $-413.5573681645$ & $-413.5487025769$ & $-413.5934852772$ & $-413.5938061263$ \\
$\bar{E}(\varepsilon = 10^{-4})$ & $-413.5572911952$ & $-413.5484958424$ & $-413.5932502365$ & $-413.5936149191$\\
$\bar{E}(\varepsilon = 3\cdot10^{-5})$ & $-413.5572861085$ & -- & -- & -- \\
\hline
MECI ($\varepsilon \to 0$) & $-413.5572839285$ & $-413.5484728719$ & $-413.5932241209$ & $-413.5935936739$\\
\hspace{5 pt} energy gap & $3.05\cdot10^{-6}$ & $1.20\cdot10^{-5}$ & $7.08\cdot10^{-7}$ & $2.23\cdot10^{-6}$\\
\hline
Seam point & $-413.5572838626$ & $-413.5484720976$ & $-413.5932233829$ & $-413.5935927308$ \\
\hspace{5 pt} energy gap & $8.66\cdot10^{-10}$ & $8.46\cdot10^{-10}$ & $8.97\cdot10^{-10}$ & $1.71 \cdot10^{-9}$\\
\hline
$\Delta$ [$\mu E_h$] & $+0.07$ & $+0.77$ & $+0.74$ & $+0.94$ \\
\hline
\end{tabular}
}
\end{table}

The model also predicts where $\mathbf{w}$ vanishes within the gh-plane through the extrapolated MECI geometry. A single calculation at this point gave energy differences of about $10^{-9}~E_h$ for all models, which confirms the intersection (see Table \ref{tab:sup_Uracil_MECI_energies}). These in-plane CI geometries (seam points) lie within 0.12~m\AA\ of the extrapolated MECI geometry for all the MECIs, the geometries are available in the data repository. Their mean energy is an upper bound to the MECI energy. The seam points were used as the center of the 2D scans. Each scan spans $\pm10$ steps along $\mathbf{g}$ and $\mathbf{h}$ such that the energy resolution is the same along both directions. The window covers about twice the region in which the underlying CC or GCC states form a complex pair.\\

\newpage
\subsubsection*{Relaxation pathway in uracil}
\noindent{\textit{Intersection points, branching planes and scan plane.}}\\
The scans use XO-GCCSD/cc-pVDZ with four constrained states (S$_0$-S$_3$). We first located the S$_0$/S$_1$ and S$_1$/S$_2$ intersection points of this model (one and two projected states in GCC respectively), starting from the extrapolated two-state MECI geometries. Each search was restricted to the branching plane of the corresponding two-state MECI. In each iteration we computed the energy gap of the intersecting pair on a small stencil of geometries displaced along $\mathbf{g}$ and $\mathbf{h}$ around the current estimate. We then fitted a conical model to these gaps and placed the next estimate at the zero of the fit. The searches were stopped once the gap fell below $2\cdot10^{-8}$~E$_\mathrm{h}$.

The scan plane is parametrized in internal coordinates $\mathbf{I}$ (a Z-matrix of bond lengths, angles and dihedral angles) as
\begin{equation}
  \mathbf{I}(\lambda, x) = \mathbf{I}_1 + \lambda\,(\mathbf{I}_2 - \mathbf{I}_1)
  + x\,\big[(1-\lambda)\,\mathbf{e}_1 + \lambda\,\mathbf{e}_2\big],
\end{equation}
where $\mathbf{I}_1$ and $\mathbf{I}_2$ are the S$_0$/S$_1$ and S$_1$/S$_2$ intersection points. Differences in dihedral angles are limited to $(-\pi, \pi]$. The parameter $\lambda$ runs along the line through both intersection points, with the S$_0$/S$_1$ point at $\lambda=0$ and the S$_1$/S$_2$ point at $\lambda=1$. The transverse vector $\mathbf{e}_i$ is constructed at each intersection point $i$. It is the unit Cartesian direction within the branching plane $\mathrm{span}(\mathbf{g}_i,\mathbf{h}_i)$ that is orthogonal to the line connecting the intersections, converted to internal coordinates (we use the branching-plane vectors of the two-state MECIs). The plane therefore contains, at each intersection, the branching-plane direction that the line itself does not cover, and both intersections are conical points within the scanned plane. The coordinate $x$ is given in \AA: at either intersection, $x$ is the Cartesian displacement along the unit vector. The sign of $\mathbf{e}_2$ is chosen so that $\mathbf{e}_1\cdot\mathbf{e}_2 > 0$. Cartesian geometries are obtained from $\mathbf{I}(\lambda,x)$ and aligned to the S$_0$/S$_1$ intersection by a Kabsch rotation (the alignment does not alter the results of the calculations, but improves restarting from neighboring geometries). The coordinate $s$ is the Cartesian arc length in \AA{} along the line $x = 0$, measured from the S$_0$/S$_1$ intersection. The S$_1$/S$_2$ intersection lies at $s = 1.580$~\AA. Around each intersection we additionally computed a cut along $x = 0$ with 21 points, spanning $\pm 0.0074$~\AA{} in $s$ around the S$_0$/S$_1$ intersection and $\pm 0.0081$~\AA{} around the S$_1$/S$_2$ intersection (steps of $7.4\cdot10^{-4}$ and $8.1\cdot10^{-4}$~\AA). Geometries for the 2D scan and 1D cuts are available in the data repository.\\

\noindent{\textit{Switching between one and two projected states in GCCSD.}}\\
In most of the scanned plane, the underlying GCCSD calculation projects one excited state (S$_1$). As S$_1$ approaches the S$_1$/S$_2$ intersection, it mixes strongly with S$_2$ and its projection becomes ill defined. In this region, S$_1$ and S$_2$ were both projected instead. The projector then spans both states and no longer depends on how they mix. Two states were projected at every grid point where the bare GCCSD S$_1$-S$_2$ gap, obtained from the calculation with one projected state, is below $c = 1$~eV. They were also projected at the three points where the calculation with one projected state did not converge, all of which lie close to the S$_1$/S$_2$ intersection. This gives 135 of the 325 grid points, covering the region around the S$_1$/S$_2$ intersection. The calculation with two projected states converged at all 135 points. In both cases the XO-CC correction constrains the same four states (S$_0$-S$_3$), and only the underlying GCCSD model changes.

\begin{figure}[hp!]
    \centering
    \includegraphics[width=\linewidth]{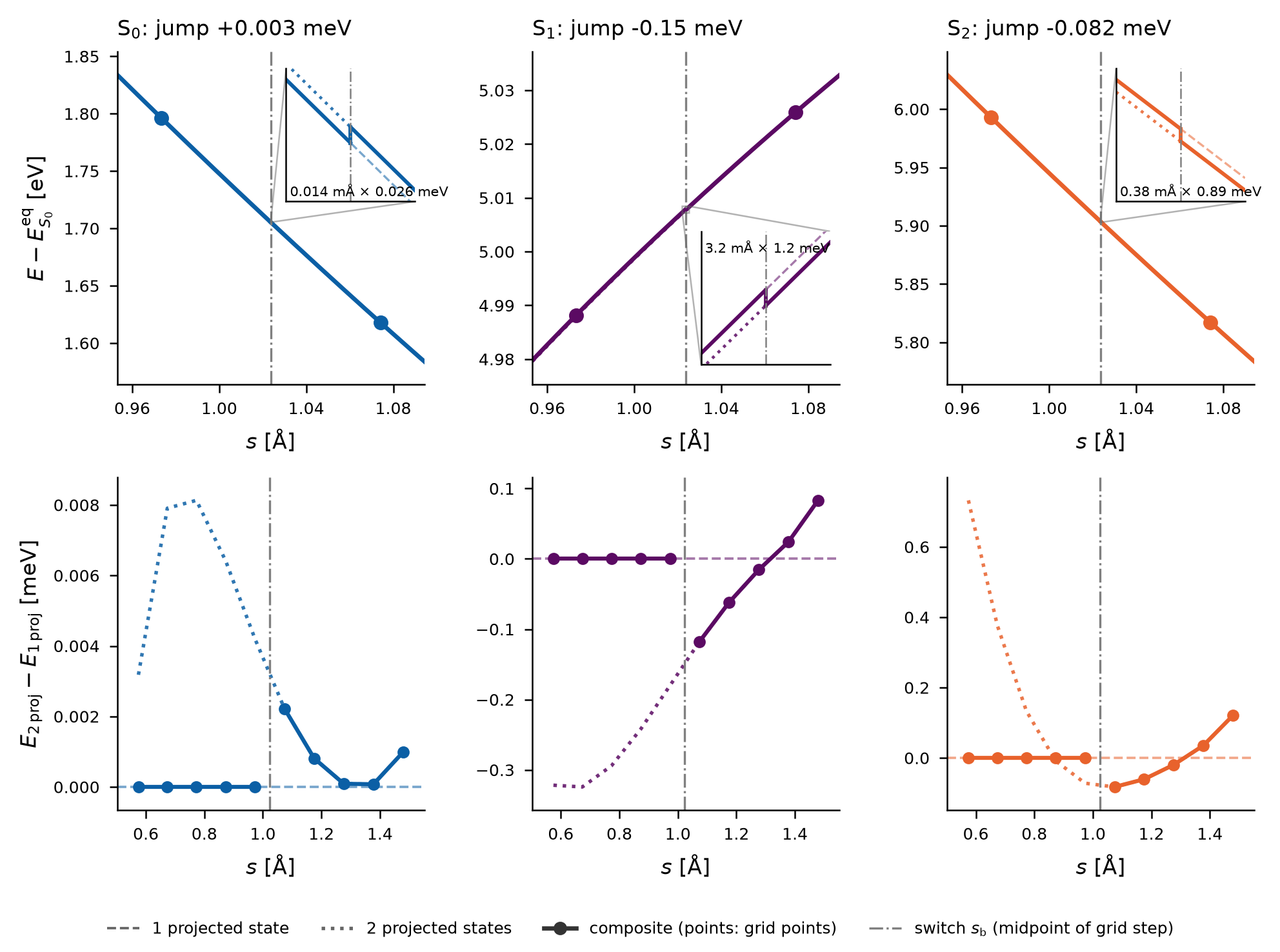}
    \caption{\footnotesize \textbf{Discontinuity of the uracil XO-GCCSD/cc-pVDZ energies at the switch from one to two projected states in the underlying GCCSD calculation, along the cut $x = 0$.} Two states are projected where the bare GCCSD S$_1$-S$_2$ gap of the calculation with one projected state is below $1$~eV. The switch is drawn at the midpoint between these two grid points (dash-dotted line). Top: S$_0$, S$_1$ and S$_2$ energies relative to the S$_0$ energy at the CCSD equilibrium geometry, around the switch. Dashed lines: one projected state; dotted lines: two projected states; solid lines: composite surface, with the neighboring grid points marked. Curves are cubic splines through the grid points. The insets magnify the region around the switch, with the window size (in m\AA{} and meV) given in each inset. Bottom: energy difference between two and one projected states along the cut; the composite surface (solid lines with markers) switches from zero to this difference at the boundary.}
    \label{fig:np12}
\end{figure}

\begin{figure}[hp!]
    \centering
    \includegraphics[width=\linewidth]{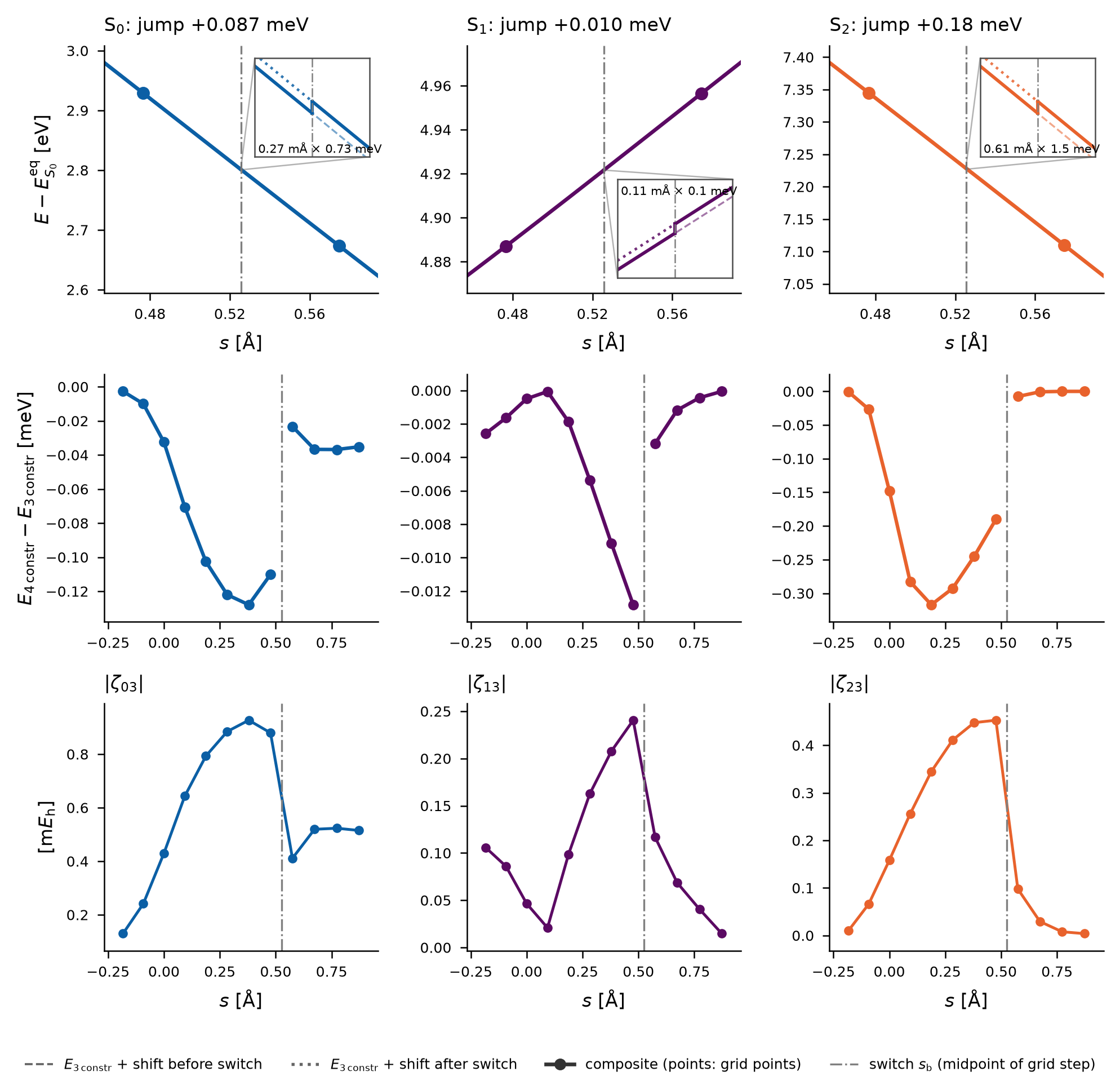}
    \caption{\footnotesize \textbf{Discontinuity of the uracil XO-GCCSD/cc-pVDZ energies where the constrained S$_3$ exchanges character with the unconstrained S$_4$, along the cut $x = 0.1$~\AA.} Bottom: couplings $|\zeta_{i3}|$ of S$_0$, S$_1$ and S$_2$ to S$_3$. Middle: effect of constraining S$_3$, $E_{4\,\mathrm{constr}} - E_{3\,\mathrm{constr}}$, where $E_{3\,\mathrm{constr}}$ is obtained with only S$_0$-S$_2$ constrained from the same GCCSD solution. It is at most 0.13~meV (S$_0$), 0.01~meV (S$_1$) and 0.32~meV (S$_2$) along the cut. Top: S$_0$, S$_1$ and S$_2$ energies relative to the S$_0$ energy at the CCSD equilibrium geometry, around the exchange. Dashed and dotted lines: the three-constrained-state energies shifted by the constant offset $E_{4\,\mathrm{constr}} - E_{3\,\mathrm{constr}}$ taken at the grid point before and after the exchange, respectively; solid lines: XO-GCCSD surface, with the neighboring grid points marked. The insets magnify the region around the exchange, with the window size given in each
    inset.} \label{fig:zeta}
\end{figure}

To quantify the discontinuity introduced by the switch, we compare the two GCCSD variants at the points of the two-projected-state region that have a neighbor outside it (15 points) and at those outside neighbors (14 points). Both calculations converged at all 29 points. Here, switching from one to two projected states changes the XO-GCCSD energies by at most 0.04~meV (S$_0$), 1.3~meV (S$_1$) and 0.9~meV (S$_2$). The bare S$_1$-S$_2$ gap is 0.79-1.00~eV at the region points and 1.00-1.26~eV at the outside neighbors, so the switch takes place where the additionally projected state is well separated from S$_1$. We show the jumps along the line $x = 0$ in Figure~\ref{fig:np12}.\\

\noindent{\textit{Effect of the S$_3$/S$_4$ discontinuity on S$_0$-S$_2$.}}\\
Where S$_3$ and S$_4$ cross,  S$_3$ changes character, and the couplings $\zeta_{i3}$ jump (Fig.~\ref{fig:Uracil_threestate} C). The resulting change in the energy of a lower-lying state S$_i$ is of second order in the coupling, $|\Delta E_i| \approx \zeta_{i3}^2/|E_i - E_3|$, and is therefore small whenever the coupling is small compared to the energy gap to S$_3$. To quantify the effect directly, we repeated the XO step at all 50 grid points adjacent to the discontinuity with only S$_0$-S$_2$ constrained, keeping the GCCSD solution fixed. The largest shifts of S$_2$ occur where S$_2$ and S$_3$ are close in energy. Across the discontinuity, the largest change in this shift between neighboring grid points (45 pairs) is 0.12, 0.07 and 0.24~meV for S$_0$, S$_1$ and S$_2$. The discontinuity therefore shifts the energies of S$_0$-S$_2$ by well below 0.5~meV. We show the jumps along the line $x = 0.1$Å in Figure~\ref{fig:zeta}. The second-order estimate of the XO-CC pairwise energy shifts reproduces the exact jumps to within 0.004~meV.

\newpage
\subsubsection*{Malonaldehyde}

All malonaldehyde calculations used XO-GCC2/6-31G* with two projected states in GCC2. S$_0$, S$_1$ and S$_2$ were the constrained states. The XO-CC residual was $10^{-7}$.

In a Hermitian-like theory (such as XO-CC theory), three states are degenerate only where the traceless part of their $3\times3$ representation vanishes. This gives five conditions, so that the degeneracy is lifted in a five-dimensional branching space. Close to the intersection, the five conditions depend approximately linearly on the nuclear displacements, allowing the determination of the intersection point by linear extrapolation: the three XO-GCC2 energies, relative to their mean, are fitted as the eigenvalues of a $3\times3$ matrix that depends linearly on the displacements. A step is then taken towards the zero of this model, and the procedure is repeated until the intersection is located. Since five conditions are to be fulfilled, five independent displacement directions are sufficient. As these, we used the space spanned by the CASSCF g- and h-vectors (rigid translations and rotations were removed) of the three two-state pairs at the three-state intersection reported by Coe and Martínez \cite{coe2005competitive}. The search started from their CASSCF intersection geometry, and the converged geometry is provided in the data repository. This geometry is a point on the three-state intersection seam, but not its minimum-energy point.

The same linear model is used to determine the plane of the 2D scan. We choose the plane on which one eigenvalue of the $3 \times 3$ matrix remains zero, so that the middle surface stays at the average energy of the three states, while the lower and upper surfaces form a double cone around it. Such a plane exists for every choice of state that belongs to the zero eigenvalue; we choose the state for which the degeneracy of the remaining two states is lifted most strongly. For malonaldehyde, this plane lifts the degeneracy equally strongly in all directions, giving a circular cone. The plane is spanned by the orthonormal Cartesian displacement vectors $x_1$ and $x_2$, both provided in the data repository. We stress that these displacement vectors are not g- and h-vectors of the XO-GCC2 model. Because the plane passes through the intersection, the three states are degenerate at its origin. A point in the scan has the geometry $\mathbf{R} = \mathbf{R}_0 + \Delta x_1\,\mathbf{x}_1 + \Delta x_2\,\mathbf{x}_2$, $\mathbf{R}_0$ being the three-state intersection geometry.

%%%%%%%%%%%%%%%% SUPPLEMENTARY FIGURES %%%%%%%%%%%%%%%

%%%%%%%%%%%%%%%% SUPPLEMENTARY TABLES %%%%%%%%%%%%%%%

%%%%%%%%%%%%%%%% SUPPLEMENTARY REFERENCES %%%%%%%%%%%%%%%

% Do NOT include a reference list in the supplement.
% All references must be in a single list at the end of the main text.
% The copyeditors will ensure that the correct reference list appears with each version of the paper
% (print, HTML, PDF, mobile app, metadata for bibliographic databases etc.)

\end{document}